\documentclass[twocolumn,prb,aps,longbibliography,10pt]{revtex4-1}

\usepackage{amsfonts,amssymb}
\usepackage[sumlimits,intlimits]{amsmath}
\usepackage{graphics}
\usepackage{graphicx}
\usepackage{color}
\usepackage{mathrsfs}
\usepackage{textcomp}
\usepackage{verbatim}
\usepackage{hyperref}    
\usepackage{bm}          

\newcommand{\be}{\begin{equation}}
\newcommand{\ee}{\end{equation}}
\newcommand{\e}{\varepsilon}

\newcommand{\Ef}{E_\text{F}}
\newcommand{\kb}{k_\text{B}}
\newcommand{\lb}{\ell_B}
\newcommand{\mub}{\mu_\text{B}}
\newcommand{\ab}{a_\text{B}^*}
\newcommand{\pf}{\textrm{PF}}

\def\bB{{\bf B}}

\def\bE{{\bf E}}

\def\bJ{{\bf J}}

\def\bsigma{{\boldsymbol \sigma}}
\def\bkappa{{\boldsymbol \kappa}}
\newcommand{\rhat}{\hat{\rho}}
\newcommand{\khat}{\hat{\kappa}}
\newcommand{\shat}{\hat{\sigma}}
\newcommand{\ahat}{\hat{\alpha}}
\newcommand{\Phat}{\hat{\Pi}}
\def\bB{{\bf B}}
\def\bE{{\bf E}}
\def\bJ{{\bf J}}
\newcommand{\ta}{\widetilde{\alpha}}
\newcommand{\tK}{\widetilde{K}}
\newcommand{\tR}{\widetilde{R}}
\newcommand{\tZ}{\widetilde{Z}}

\begin{document}

\title{Large, nonsaturating thermopower in a quantizing magnetic field}

\author{Brian Skinner}
\author{Liang Fu}
\affiliation{Massachusetts Institute of Technology, Cambridge, MA 02139 USA}

\date{\today}

\begin{abstract}

The thermoelectric effect is the generation of an electrical voltage from a temperature gradient in a solid material due to the diffusion of free charge carriers from hot to cold.  Identifying materials with large thermoelectric response is crucial for the development of novel electric generators and coolers.  In this paper we consider theoretically the thermopower of Dirac/Weyl semimetals subjected to a quantizing magnetic field.  We contrast their thermoelectric properties with those of traditional heavily-doped semiconductors and we show that, under a sufficiently large magnetic field, the thermopower of Dirac/Weyl semimetals grows linearly with the field without saturation and can reach extremely high values.  Our results suggest an immediate pathway for achieving record-high thermopower and thermoelectric figure of merit, and they compare well with a recent experiment on Pb$_{1-x}$Sn$_x$Se.

\end{abstract}

\maketitle

\section{Introduction}

When a temperature gradient is applied across a solid material with free electronic carriers, a voltage gradient arises as carriers migrate from the hot side to the cold side. The strength of this thermoelectric effect is characterized by the Seebeck coefficient $\alpha$, defined as the ratio between the voltage difference $\Delta V$ and the temperature difference $\Delta T$; the absolute value of $\alpha$ is referred to as the thermopower.  Finding materials with large thermopower is vital for the development of thermoelectric generators and thermoelectric coolers -- devices which can transform waste heat into useful electric power, or electric current into cooling power.\cite{ioffe_semiconductor_1957, dresselhaus_new_2007, shakouri_recent_2011}

The effectiveness of a thermoelectric material for power applications is quantified by its thermoelectric figure of merit
\be
Z T = \alpha^2 \sigma  T/\kappa,
\label{eq:ZT}
\ee
where $\sigma$ is the electrical conductivity, $T$ is the temperature, and $\kappa$ is the thermal conductivity.  To design a material with large thermoelectric figure of merit, one can try in general to use either an insulator, such as an intrinsic or lightly-doped semiconductor, or a metal, such as a heavily-doped semiconductor.  In an insulator the thermopower can be large, of order $E_0/(e T)$, where $e$ is the electron charge and $E_0$ is the difference in energy between the chemical potential and the nearest band mobility edge.\cite{Chen_anomalously_2013}  However, obtaining such a large thermopower comes at the expense of an exponentially small, thermally-activated conductivity, $\sigma \propto \exp(-E_0/\kb T)$, where $\kb$ is the Boltzmann constant.  Since the thermal conductivity in general retains a power-law dependence on temperature due to phonons, the figure of merit $ZT$ for insulators is typically optimized when $E_0$ and $\kb T$ are of the same order of magnitude.  This yields a value of $Z T$ that can be of order unity but no larger.\cite{Mahan_figure_1989}

On the other hand, metals have a robust conductivity, but usually only a small Seebeck coefficient $\alpha$.  In particular, in the best-case scenario where the thermal conductivity due to phonons is much smaller than that of electrons, the Wiedemann-Franz law dictates that the quantity $\sigma T/\kappa$ is a constant of order $(e/\kb)^2$.  The Seebeck coefficient, however, is relatively small in metals, of order $\kb^2T/(e \Ef)$, where $\Ef \gg \kb T$ is the metal's Fermi energy.  If the temperature is increased to the point that $\kb T > \Ef$, the Seebeck coefficient typically saturates at a constant of order $\kb/e$.  The maximum value of the figure of merit in metals is therefore obtained when $\kb T$ is of the same order as $\Ef$, and again ones arrives at an apparent maximum value of $Z T$ that is of order unity at best.

In this paper we show that these limitations can be circumvented by considering the behavior of doped nodal semimetals in a strong magnetic field, for which $Z T \gg 1$ is in fact possible.
Crucial to our proposal is a confluence of three effects.  First, a sufficiently high magnetic field produces a large enhancement of the electronic density of states and a reduction in the Fermi energy $\Ef$.  Second, a quantizing magnetic field assures that the transverse $\bE \times \bB$ drift of carriers plays a dominant role in the charge transport, and this allows both electrons and holes to contribute \emph{additively} to the thermopower, rather than subtractively as in the zero-field situation.  Third, in materials with a small band gap and electron-hole symmetry, the Fermi level remains close to the band edge in the limit of large magnetic field, and this allows the number of thermally-excited electrons and holes to grow with magnetic field even while their difference remains fixed. These three effects together allow the thermopower to grow without saturation as a function of magnetic field. 

\section{Relation Between Seebeck Coefficient and Entropy}
 
The Seebeck coefficient is usually associated, conceptually, with the entropy per charge carrier.  
%
In a large magnetic field, and in a generic system with some concentrations $n_e$ of electrons and $n_h$ of holes, the precise relation between carrier entropy and thermopower can be derived using the following argument. Let the magnetic field $\bB$ be oriented in the $z$ direction, and suppose that an electric field $\bE$ is directed along the $y$ direction. Suppose also that the magnetic field is strong enough that $\omega_c \tau \gg 1$, where $\omega_c$ is the cyclotron frequency and $\tau$ is the momentum scattering time, so that carriers complete many cyclotron orbits without scattering.  In this situation charge carriers acquire an $\bE \times \bB$ drift velocity in the $x$ direction, with magnitude $v_d = E/B$.  Importantly, the direction of drift is identical for both negatively charged electrons and positively charged holes, so that drifting electrons and holes contribute additively to the heat current but oppositely to the electrical current.  This situation is illustrated in Fig.\ 1.

\begin{figure}[b]
	\centering
	\includegraphics[width=0.4\textwidth]{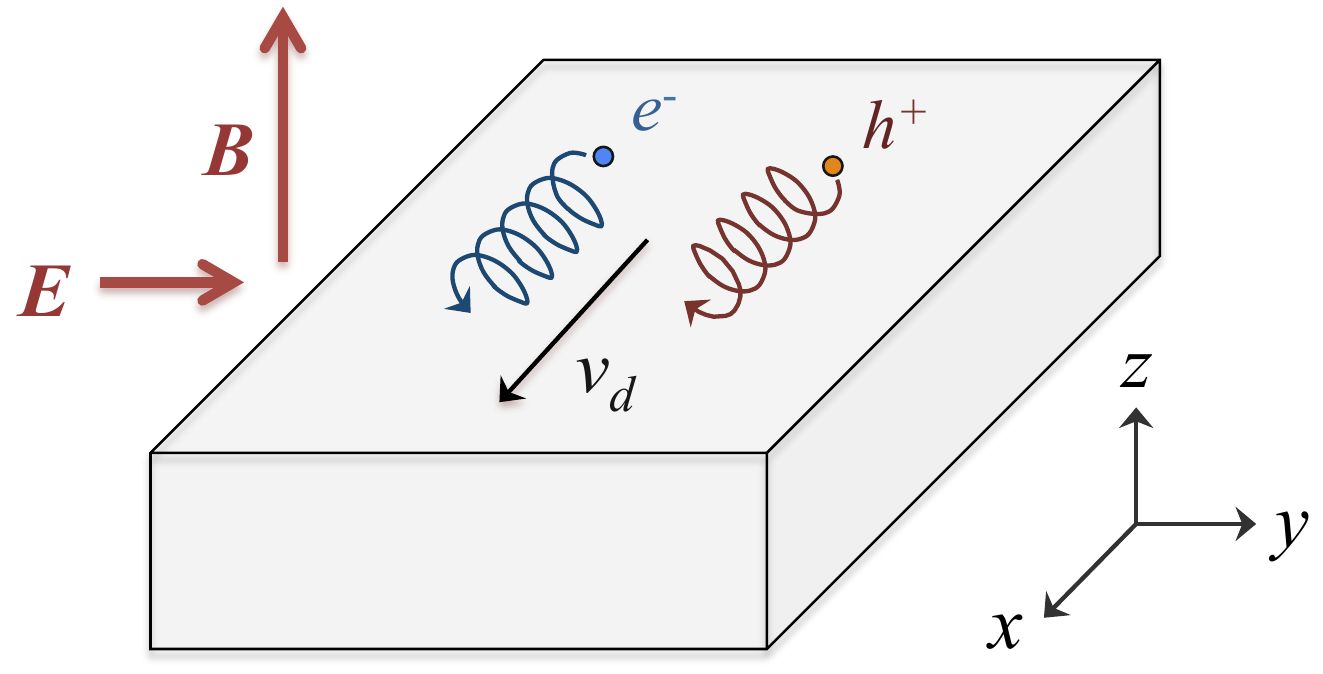}
	\caption{Schematic depiction of the $\bE \times \bB$ drift of carriers in a strong magnetic field. Electrons (labeled $e^{-}$) and holes (labeled $h^{+}$) drift in the same direction under the influence of crossed electric and magnetic fields. Both signs of carrier contribute additively to the heat current in the $y$ direction and subtractively to the electric current in the $y$ direction, which leads to a large Peltier heat $\Pi_{xx}$ and therefore to a large thermopower $\alpha_{xx}$.}
	\label{fig:drifting}
\end{figure}

To understand the Seebeck coefficient $\alpha_{xx}$ in the $x$ direction, one can exploit the Onsager symmetry relation between the coefficients $\alpha_{ij}$ of the thermoelectric tensor and the coefficients $\Pi_{ij}$ of the Peltier heat tensor: $\alpha_{ij}(B) = \Pi_{ji}(-B)/T$.  The Peltier heat is defined by $J^Q_{i} = \Pi_{ij} J^e_{j}$, where $\bJ^Q$ is the heat current density at a fixed temperature and $\bJ^e$ is the electrical current density.  In the setup we are considering, the electrical current in the $x$ direction is given simply by $J^e_x = e v_d (n_h - n_e)$.

In sufficiently large magnetic fields, the flow of carriers in the $x$ direction is essentially dissipationless. In this case the heat current in the $x$ direction is related to the \emph{entropy current} $J^s_x$ by the law governing reversible processes: $J^Q_x = T J^s_x$.  This relation is valid in general when the the Hall conductivity $\sigma_{xy}$ is much larger in magnitude than the longitudinal  conductivity $\sigma_{xx}$; for a system with only a single sign of carriers this condition is met when $\omega_c \tau \gg 1$.  If we define $s_e$ and $s_h$ as the entropy per electron and per hole, respectively, then $J^s_x = v_d(n_e s_e + n_h s_h)$, since electrons and holes both drift in the $x$ direction.  Putting these relations together, we arrive at a Seebeck coefficient $\alpha_{xx} = \Pi_{xx}/T = (J^Q_x)/(T J^e_x)$ that is given by
\be
\alpha_{xx} = \frac{n_h s_h + n_e s_e}{e(n_h - n_e)} \equiv \frac{S}{e n}.
\label{eq:alphax}
\ee
In other words, the Seebeck coefficient in the $x$ direction is given simply by the \emph{total} entropy density $S$ divided by the \emph{net} carrier charge density $e n$.  This relation between entropy and thermopower in a large transverse magnetic field has been recognized for over fifty years and explained by a number of authors \cite{obraztsov_thermal_1965, tsendin_theory_1966, jay-gerin_thermoelectric_1974, abrikosov_fundamentals_1988, bergman_theory_2010}, but it is usually applied only to systems with one sign of carriers.  As we show below, it has dramatic implications for the thermopower in gapless three-dimensional (3D) semimetals, where both electrons and holes can proliferate at small $\Ef \ll \kb T$.  

In the remainder of this paper we focus primarily on the thermopower $\alpha_{xx}$ in the directions transverse to the magnetic field, which can be described simply according to Eq.\ (\ref{eq:alphax}).  At the end of the paper we comment briefly on the thermopower along the direction of the magnetic field, which has less dramatic behavior and which saturates in all cases at $\sim (\kb/e)$ in the limit of large magnetic field.
We also neglect everywhere the contribution to the thermopower arising from phonon drag.  This is valid provided that the temperature and Fermi energy $\Ef$ are low enough that $(\kb T/\Ef) \gg (T/\Theta_D)^3$, where $\Theta_D$ is the Debye temperature. \cite{ziman_principles_1972}
Such low-temperature and low-$\Ef$ systems are the focus of this paper (although it should be noted that phonon drag tends to \emph{increase} the thermopower \cite{jay-gerin_thermoelectric_1975}).

When the response coefficients governing the flow of electric and thermal currents have finite transverse components, as introduced by the magnetic field, the definition of the figure of merit $ZT$ should be generalized from the standard expression of Eq.\ (\ref{eq:ZT}).  This generalized definition can be arrived at by considering the thermodynamic efficiency of a thermoelectric generator with generic thermoelectric, thermal conductivity, and resistivity tensors.  The resulting generalized figure of merit is derived in Appendix \ref{app:ZBT}, and is given by
\be 
Z_B T = \frac{\alpha_{xx}^2 T}{\kappa_{xx} \rho_{xx}} \frac{\left(1 - \frac{\alpha_{xy}}{\alpha_{xx}} \frac{\kappa_{xy}}{\kappa_{xx}}\right)^2}{\left(1 + \frac{\kappa_{xy}^2}{\kappa_{xx}^2}\right) \left(1 - \frac{\alpha_{xy}^2 T}{\kappa_{xx} \rho_{xx}}\right) },
\label{eq:ZBT}
\ee
where $\rho_{xx}$ is the longitudinal resistivity.  Similarly, the thermoelectric power factor, which determines the maximal electrical power that can be extracted for a given temperature difference, is given by
\be 
\pf = \frac{\alpha_{xx}^2}{\rho_{xx}} \frac{\left(1 - \frac{\alpha_{xy}}{\alpha_{xx}} \frac{\kappa_{xy}}{\kappa_{xx}}\right)^2}{1 - \frac{\alpha_{xy}^2 T}{\kappa_{xx} \rho_{xx}} } .
\ee
In the limit of $\omega_c \tau \gg 1$ that we are considering, $\alpha_{xy} \ll \alpha_{xx}$, and therefore for the remainder of this paper we restrict our analysis to the case $\alpha_{xy} = 0$. 

In situations where phonons do not contribute significantly to the thermal conductivity, we can simplify Eq.\ (\ref{eq:ZBT}) by exploiting the Wiedemann-Franz relation, $\bkappa = c_0 (\kb/e)^2 T \bsigma$, where $c_0$ is a numeric coefficient of order unity and $\bkappa$ and $\bsigma$ represent the full thermal conductivity and electrical conductivity tensors. This relation remains valid even in the limit of large magnetic field, so long as electrons and holes are good quasiparticles.\cite{abrikosov_fundamentals_1988}  In the limit of strongly degenerate statistics, where either $\Ef \gg \kb T$ or the band structure has no gap, $c_0$ is given by the usual value $c_0 = \pi^2/3$ corresponding to the Lorentz ratio.  In the limit of classical, nondegenerate statistics, where $\Ef \ll \kb T$ and the Fermi level resides inside a band gap, $c_0$ takes the value corresponding to classical thermal conductivity: $c_0 = 4/\pi$.
Inserting the Wiedemann-Franz relation into Eq.\ (\ref{eq:ZBT}) and setting $\alpha_{xy} = 0$ gives
\be
Z_B T = \frac{\alpha_{xx}^2}{c_0 (\kb/e)^2}.
\label{eq:ZBTWF}
\ee
In other words, when the phonon conductivity is negligible the thermoelectric figure of merit is given to within a multiplicative constant by the square of the Seebeck coefficient, normalized by its natural unit $\kb/e$.  As we show below, in a nodal semimetal $\alpha_{xx}/(\kb/e)$ can be parametrically large under the influence of a strong magnetic field, and thus the figure of merit $Z_B T$ can far exceed the typical bound for heavily-doped semiconductors.

In situations where phonons provide a dominant contribution to the thermal conductivity, so that the Wiedemann-Franz law is strongly violated, one generically has $\kappa_{xx} \gg \kappa_{xy}$, and Eq.\ (\ref{eq:ZBT}) becomes 
\be 
Z_B T = \frac{\alpha_{xx}^2 T}{\kappa_{xx} \rho_{xx}}.
\label{eq:ZBTphonons}
\ee

\section{Heavily-Doped Semiconductors}

In this section we present a calculation of the thermopower $\alpha_{xx}$ for a heavily-doped semiconductor, assuming for simplicity an isotropic band mass $m$ and a fixed carrier concentration $n$.  (In other words, we assume sufficiently high doping that carriers are not localized onto donor/acceptor impurities by magnetic freezeout.\cite{pepper_metal-insulator_1979})  This classic problem has been considered in various limiting cases by previous authors.\cite{obraztsov_thermal_1965, jay-gerin_thermoelectric_1974, jay-gerin_thermoelectric_1975, arora_thermoelectric_1979} Here we briefly present a general calculation and recapitulate the various limiting cases, both for the purpose of conceptual clarity and to provide contrast with the semimetal case.

Full details of the thermopower calculation at arbitrary $B$ and $T$ are presented in Appendix \ref{app:semiconductor}, and an example of this calculation is shown in Fig.\ 2.  This plot considers a temperature $T \ll \Ef^{(0)}/\kb$, where $\Ef^{(0)}$ is the Fermi energy at zero magnetic field.  The asymptotic behaviors evidenced in this figure can be understood as follows.

In the limit of vanishing temperature, the chemical potential $\mu$ is equal to the Fermi energy $\Ef$, and the entropy per unit volume
\be
S \simeq \frac{\pi^2}{3} \kb^2 T \nu(\mu),
\label{eq:SDOS}
\ee
where $\nu(\mu)$ is the density of states at the Fermi level.
At weak enough magnetic field that $\hbar \omega_c \ll \Ef$, the density of states is similar to that the usual 3D electron gas, and the corresponding thermopower is
\be
\alpha_{xx} \simeq \frac{\kb}{e} \left( \frac{\pi}{3} N_v \right)^{2/3} \frac{\kb T m}{\hbar^2 n^{2/3}},
\label{eq:alphasemismallB}
\ee
where $N_v$ is the degeneracy per spin state (the valley degeneracy) and $\hbar$ is the reduced Planck constant.
As the magnetic field is increased, the density of states undergoes quantum oscillations that are periodic in $1/B$, which are associated with individual Landau levels passing through the Fermi level.  These oscillations are reflected in the thermopower, as shown in Fig.\ 2.

\begin{figure}[b]
	\centering
	\includegraphics[width=0.45\textwidth]{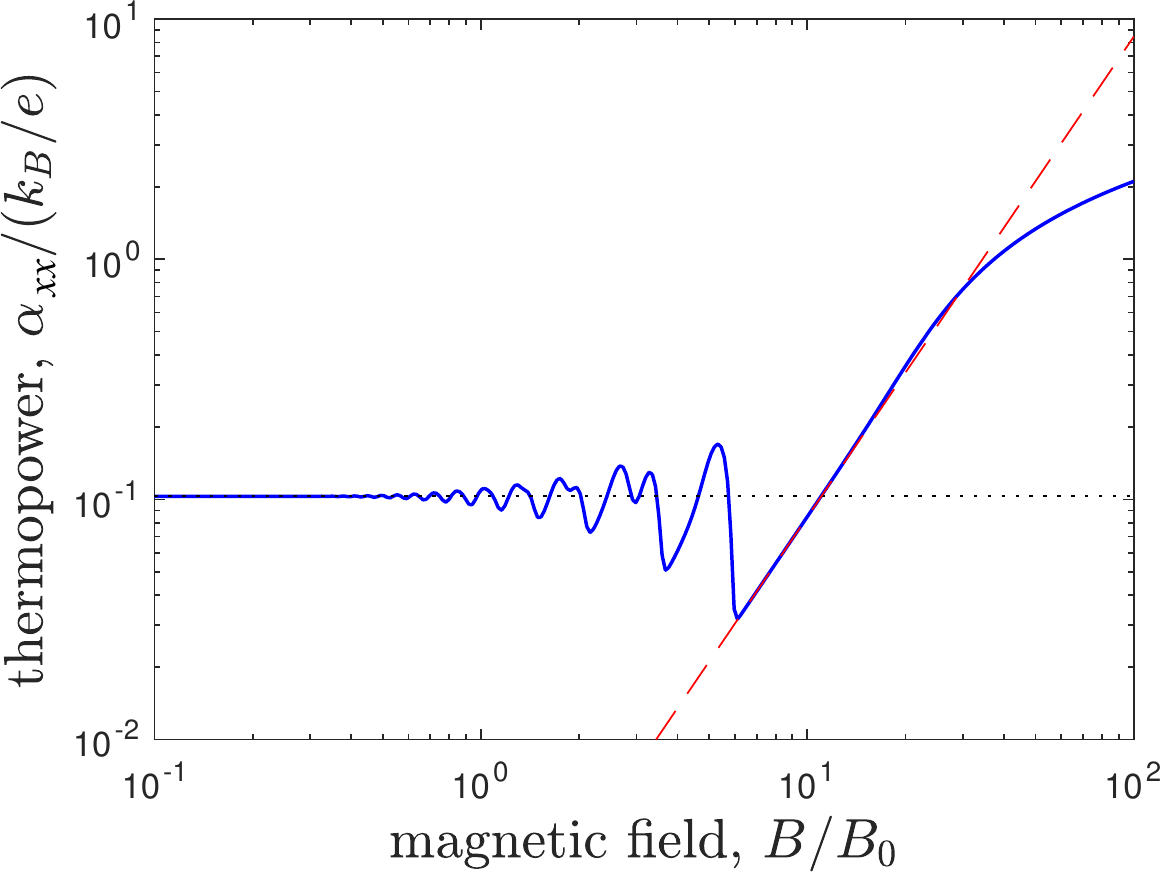}
	\caption{Thermopower in the transverse direction, $\alpha_{xx}$, as a function of magnetic field for a degenerate semiconductor with parabolic dispersion relation.   The magnetic field is plotted in units of $B_0 = \hbar n^{2/3}/e$.  The temperature is taken to be $T = 0.02 \Ef^{(0)}/\kb$, and for simplicity we have set $N_v = 1$ and $g = 2$. 
The dotted line shows the limiting result of Eq.\ (\ref{eq:alphasemismallB}) for small $B$, and the dashed line shows the result of Eq.\ (\ref{eq:EQLsemi}) for the extreme quantum limit.  At very large magnetic field the thermopower saturates at $\sim \kb/e$, with only a logarithmic dependence on $B$ and $T$, as suggested by Eq.\ (\ref{eq:alphaclassicalsemi})}
	\label{fig:semi}
\end{figure}

Of course, 
Eq.\ (\ref{eq:alphasemismallB}) assumes that impurity scattering is sufficiently weak that $\omega_c \tau \gg 1$.  For the case of a doped and uncompensated semiconductor where the scattering rate is dominated by elastic collisions with donor/acceptor impurities, this limit corresponds to\cite{dingle_scattering_1955} $\lb \ll \ab$, where $\lb = \sqrt{\hbar/eB}$ is the magnetic length and $\ab = 4 \pi \epsilon \hbar^2/(m e^2)$ is the effective Bohr radius, with $\epsilon$ the permittivity.  In the opposite limit of small $\omega_c \tau$, the thermopower at $\kb T \ll \Ef$ is given by the Mott formula \cite{abrikosov_fundamentals_1988}
\be
\alpha = \frac{\kb}{e} \frac{\pi^2}{3} \frac{1}{\sigma} \left. \left( \frac{d \sigma(E)}{d E} \right) \right|_{E = \mu},  \hspace{5mm}  (\textrm{at } B = 0),
\label{eq:B0alpha}
\ee
where $\sigma(E)$ is the low-temperature conductivity of a system with Fermi energy $E$.
In a doped semiconductor with charged impurity scattering, the conductivity $\sigma \propto \Ef^3$, and Eq.\ (\ref{eq:B0alpha}) gives a value that is twice larger than that of Eq.\ (\ref{eq:alphasemismallB}).

When the magnetic field is made so large that $\hbar \omega_c \gg \Ef$, electrons occupy only the lowest Landau level and the system enters the extreme quantum limit.  At such high magnetic fields the density of states rises strongly with increased $B$, as more and more flux quanta are threaded through the system and more electron states are made available at low energy.  As a consequence, the Fermi energy falls relative to the energy of the lowest Landau level, and $\Ef$ and $\nu(\mu)$ are given by
\begin{eqnarray}
\Ef(B) -\frac{\hbar \omega_c}{2} & = & \frac{2 \pi^4 \hbar^2 n^2 \lb^4}{m N_s^2 N_v^2} \propto 1/B^2
\nonumber \\
\nu(\mu) & = & \frac{m N_s^2 N_v^2}{4 \pi^4 \hbar^2 n \lb^4} \propto B^2.
\end{eqnarray}
Here $N_s$ denotes the spin degeneracy at high magnetic field; $N_s = 1$ if the lowest Landau level is spin split by the magnetic field and $N_s = 2$ otherwise.  So long as the thermal energy $\kb T$ remains smaller than $\Ef$, Eq.\ (\ref{eq:SDOS}) gives a thermopower
\be
\alpha_{xx} = \frac{\kb}{e} \frac{N_s^2 N_v^2}{12 \pi^2}  \frac{m e^2 B^2 \kb T}{\hbar^4 n}.
\label{eq:EQLsemi}
\ee

Finally, if the magnetic field is so large that $\kb T$ becomes much larger than the zero-temperature Fermi energy, then the distribution of electron momenta $p$ in the field direction is well described by a classical Boltzmann distribution: $f \simeq \textrm{const.} \times \exp[-p^2/(2m \kb T)]$.  Using this distribution to calculate the entropy gives a thermopower
\be
\alpha_{xx} \simeq \frac{1}{2} \frac{\kb}{e} \ln \left( \frac{m \kb T N_v^2 N_s^2}{\hbar^2 n^2 \lb^4} \right).
\label{eq:alphaclassicalsemi}
\ee
In other words, in the limit of such large magnetic field that $\hbar \omega_c \gg \kb T \gg \Ef$, the thermopower saturates at a value $\sim \kb/e$ with only a logarithmic dependence on the magnetic field. [The argument of the logarithm in Eq.\ (\ref{eq:alphaclassicalsemi}) is proportional to $\kb T / \Ef(B)$.]  This result is reminiscent of the thermopower in non-degenerate (lightly-doped) semiconductors at high temperature,\cite{herring_transport_1955} where the thermopower becomes $\sim (\kb/e) \ln(T)$.

\section{Dirac/Weyl Semimetals}

Let us now consider the case where quasiparticles have a linear dispersion relation and no band gap (or, more generally, a band gap that is smaller than $\kb T$), as in 3D Dirac or Weyl semimetals.  Here we assume, for simplicity, that the Dirac velocity $v$ is isotropic in space, so that in the absence of magnetic field the quasiparticle energy is given simply by $\e = \pm v p$ where $p$ is the magnitude of the quasiparticle momentum.  The carrier density $n$ is constant as a function of magnetic field, since the gapless band structure precludes the possibility of magnetic freezeout of carriers.
A generic calculation of the thermopower $\alpha_{xx}$ is presented in Appendix \ref{app:Dirac}, and an example of our result is plotted in Fig.\ 3.

\begin{figure}[b]
	\centering
	\includegraphics[width=0.45\textwidth]{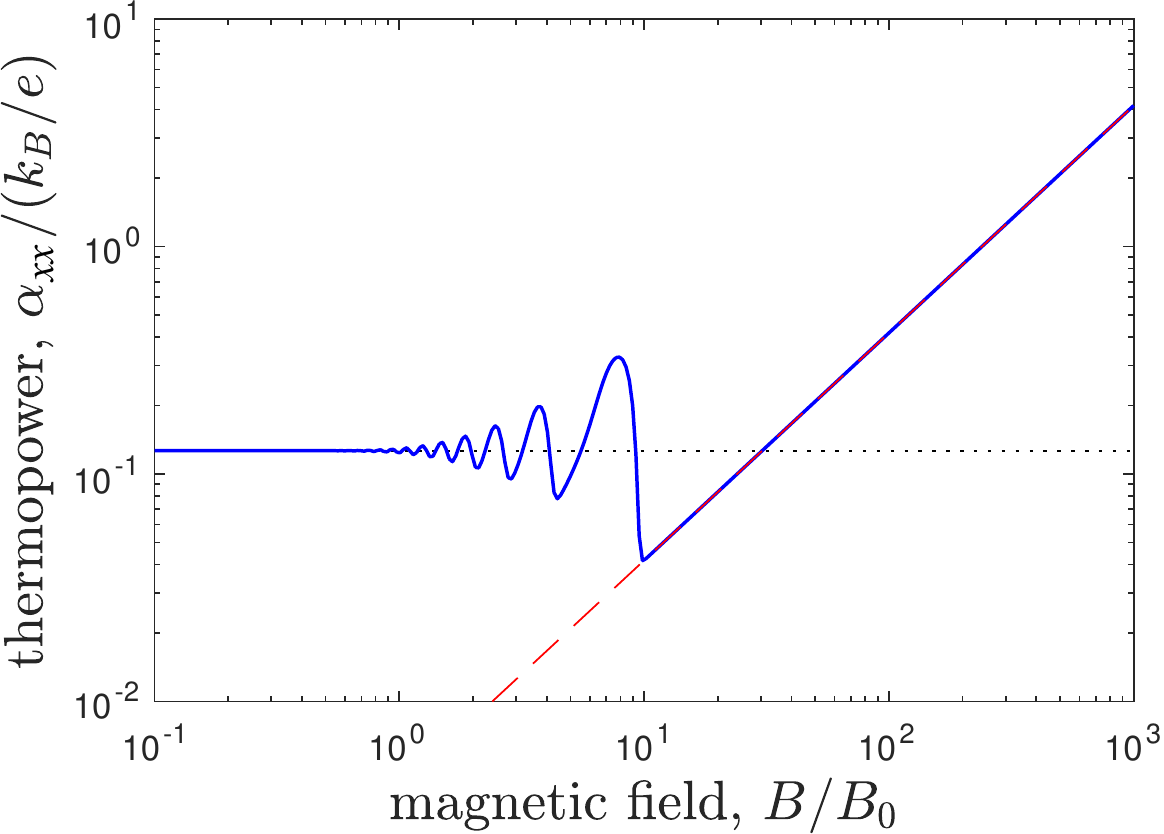}
	\caption{Thermopower in the transverse direction as a function of magnetic field for a gapless semimetal with linear dispersion relation. Units of magnetic field are $B_0 = \hbar n^{2/3}/e$. In this example the temperature is taken to be $T = 0.01 \Ef^{(0)}/\kb$ and $N_v = 1$.  The dotted line is the low-field limit given by Eq.\ (\ref{eq:alphaDiracsmallB}) and the dashed line is the extreme quantum limit result of Eq.\ (\ref{eq:alphaEQLDirac}).  Unlike the semiconductor case, at large magnetic field the thermopower continues to grow with increasing $B$ without saturation.
	}
	\label{fig:Dirac}
\end{figure}

The limiting cases for the thermopower can be understood as follows.  In the weak field regime $\hbar \omega_c \ll \Ef$, the electronic density of states is relatively unmodified by the magnetic field, and one can use Eq.\ (\ref{eq:SDOS}) with the zero-field density of states $\nu(\mu) = (9 N_v/ \pi^2)^{1/3} n^{2/3}/\hbar v$.  This procedure gives a thermopower
\be
\alpha_{xx} \simeq \frac{\kb}{e} \left( \frac{\pi^4}{3} \right)^{1/3} \frac{\kb T}{\hbar v} \left( \frac{N_v}{n} \right)^{1/3}.
\label{eq:alphaDiracsmallB}
\ee
Here $N_v$ is understood as the number of Dirac nodes; for a Weyl semimetal, $N_v$ is equal to half the number of Weyl nodes.
Equation (\ref{eq:alphaDiracsmallB}) applies only when $\omega_c \tau \gg 1$.  If the dominant source of scattering comes from uncompensated donor/acceptor impurities,\cite{skinner_coulomb_2014} then the condition $\omega_c \tau \gg 1$ corresponds to $B \gg e n^{2/3}/(4 \pi \epsilon v)$.  In the opposite limit of small $\omega_c \tau$, one can evaluate the thermopower using the Mott relation [Eq.\ (\ref{eq:B0alpha})].  A Dirac material with Coulomb impurity scattering has $\sigma(E) \propto E^4$,\cite{skinner_coulomb_2014} so in the limit $\omega_c \tau \ll 1$ the thermopower is larger than Eq.\ (\ref{eq:alphaDiracsmallB}) by a factor $4/3$.

As the magnetic field is increased, the thermopower undergoes quantum oscillations as higher Landau levels are depopulated.  At a large enough field that $\hbar v /\lb > \Ef$, the system enters the extreme quantum limit and the Fermi energy and density of states become strongly magnetic field dependent.  In particular,
%
%
\begin{eqnarray}
\mu & \simeq & \frac{ 2 \pi^2}{N_v} \hbar v n \lb^2 \propto 1/B \nonumber \\
\nu(\mu) & \simeq & \frac{N_v}{2 \pi^2 \hbar v \lb^2} \propto B.
\label{eq:muDiracEQL}
\end{eqnarray}
The rising density of states implies that the thermopower also rises linearly with magnetic field.  From Eq.\ (\ref{eq:SDOS}),
\be
\alpha_{xx} \simeq \frac{\kb}{e} \frac{N_v}{6} \frac{\kb T e B}{\hbar^2 v n}.
\label{eq:alphaEQLDirac}
\ee

Remarkably, this relation does not saturate when $\mu$ becomes smaller than $\kb T$.  Instead, Eq.\ (\ref{eq:alphaEQLDirac}) continues to apply up to arbitrarily high values of $B$, as $\mu$ declines and the density of states continues to rise with increasing magnetic field.  One can think that this lack of saturation comes from the gapless band structure, which guarantees that there is no regime of temperature for which carriers can described by classical Boltzmann statistics, unlike in the semiconductor case when the chemical potential falls below the band edge. 
In more physical terms, 
the non-saturating thermopower is associated with a proliferation of electrons and holes at large $(\kb T)/\mu$.  Unlike in the case of a semiconductor with large band gap, for the Dirac/Weyl semimetal the number of electronic carriers is not fixed as a function of magnetic field.  As $\mu$ falls and the density of states rises with increasing magnetic field, the concentrations of electrons and holes both increase even as their difference $n = n_e - n_h$ remains fixed.  Since in a strong magnetic field both electrons and holes contribute additively to the thermopower (as depicted in Fig.\ 1), the thermopower $\alpha_{xx}$ increases without bound as the magnetic field is increased.  This is notably different from the usual situation of semimetals at $B = 0$, where electrons and holes contribute oppositely to the thermopower.\cite{gurevich_nature_1995}

The unbounded growth of $\alpha_{xx}$ with magnetic field also allows the figure of merit $Z_B T$ to grow, in principle, to arbitrarily large values.  For example, in situations where the Wiedemann-Franz law holds, Eq.\ (\ref{eq:ZBTWF}) implies a figure of merit that grows without bound in the extreme quantum limit as $B^2 T^3$.  On the other hand, if the phonon thermal conductivity is large enough that the Wiedemann-Franz law is violated, then the behavior of the figure of merit depends on the field and temperature dependence of the resistivity.  As we discuss below, in the common case of a mobility that declines inversely with temperature, the figure of merit grows as $B^2 T^2$, and can easily become significantly larger than unit in experimentally accessible conditions.

\section{Discussion} 

\noindent {\it Thermopower in the longitudinal direction.}

So far we have concentrated on the thermopower $\alpha_{xx}$ in the direction transverse to the magnetic field; 
let us now briefly comment on the behavior of the thermopower $\alpha_{zz}$ in the field direction.
At low temperature $\kb T \ll \Ef$ the thermopower $\alpha_{zz}$ can be estimated using the usual zero-field expression, Eq.\ (\ref{eq:B0alpha}), where $\sigma$ is understood as $\sigma_{zz}$.
%
%
This procedure gives the usual thermopower $\alpha_{zz} \sim \kb^2 T/(e \Ef)$.  Such a result has a weak dependence on magnetic field outside the extreme quantum limit, $\hbar \omega_c \ll \Ef$, and rises with magnetic field when the extreme quantum limit is reached in the same way that $\alpha_{xx}$ does.  That is, $\alpha_{zz} \propto B^2$ for the semiconductor case [as in Eq.\ (\ref{eq:EQLsemi})] and $\alpha_{zz} \propto B$ for the Dirac semimetal case [as in Eq.\ (\ref{eq:alphaEQLDirac})], provided that $\Ef \gg \kb T$.

However, when the magnetic field is made so strong that $\Ef(B) \ll \kb T$, the thermopower $\alpha_{zz}$ saturates. 
This can be seen by considering the definition of thermopower in terms of the coefficients of the Onsager matrix:
$\alpha = L^{12}/L^{11}$, where $L^{11} = -\int dE f'(E) \sigma(E)$ and $L^{12}  = -1/(eT) \int dE f'(E) (E - \mu) \sigma(E)$.\cite{ashcroft_solid_1976}  In the limit where $\kb T \gg |\mu|$, the coefficient $L^{11}$ is equal to $\sigma$ while $L^{12}$ is of order $\kb \sigma/e$.  Thus, unlike the behavior of $\alpha_{xx}$, the growth of the thermopower in the field direction saturates when $\alpha_{zz}$ becomes as large as $\sim \kb/e$.  As alluded to above, this difference arises because in the absence of a strong Lorentz force electrons and holes flow in opposite directions under the influence of an electric field and thereby contribute oppositely to the thermopower.  It is only the strong $\bE \times \bB$ drift, which works in the same direction for both electrons and holes, that allows the Dirac semimetal to have an unbounded thermopower $\alpha_{xx}$ in the perpendicular direction.

\

\noindent \textit{Experimental realizations.}


In semiconductors, achieving a thermopower of order $\kb/e$ is relatively common, particularly when
the donor/acceptor states are shallow and the doping is light.  
Nonetheless, we are unaware of any experiments that clearly demonstrate the $B^2$ enhancement of $\alpha_{xx}$ implied by Eq.\ (\ref{eq:EQLsemi}) for heavily-doped semiconductors.  Achieving this result requires a semiconductor that can remain a good conductor even at low electron concentration and low temperature, so that the extreme quantum limit is achievable at not-too-high magnetic fields.  This condition is possible only for semiconductors with relatively large effective Bohr radius $\ab$, either because of a small electron mass or a large dielectric constant.  For example, the extreme quantum limit has been reached in 3D crystals of HgCdTe,\cite{rosenbaum_magnetic-field-induced_1985} InAs,\cite{shayegan_magnetic-field-induced_1988}, and SrTiO$_3$.\cite{kozuka_vanishing_2008, bhattacharya_spatially_2016}  SrTiO$_3$, in particular, represents a good platform for observing large field enhancement of the thermopower, since its enormous dielectric constant allows one to achieve metallic conduction with extremely low Fermi energy.  For example, using the conditions of the experiments in Ref.\ \cite{bhattacharya_spatially_2016}, where $n \sim 5 \times 10^{16}$\,cm$^{-3}$ and $T = 20$\,mK, the value of $\alpha_{xx}$ can be expected to increase $\approx 50$ times between $B = 5$\,T and $B = 35$\,T.
The corresponding increase in the figure of merit is similarly large, although at such low temperatures the magnitude of $Z_B T$ remains relatively small.

More interesting is the application of our results to nodal semimetals, where $\alpha_{xx}$ does not saturate at $\sim \kb/e$, but continues to grow linearly with $B$ without saturation.  In fact, such behavior was recently seen by the authors of Ref.\ \cite{liang_evidence_2013}.  These authors measured $\alpha_{xx}$ in the Dirac material Pb$_{1-x}$Sn$_x$Se as a function of magnetic field, and observed a result strikingly similar to that of Fig.\ 3, with quantum oscillations in $\alpha_{xx}$ at low field followed by a continuous linear increase with $B$ upon entering the extreme quantum limit.  Indeed, our theoretical results for $\alpha_{xx}$ agree everywhere with their measured value to within a factor $2$ (the slight disagreement may be due to spatial anisotropy of the Dirac velocity
).  Our results suggest that the linear increase in $\alpha_{xx}$ should continue without bound as $B$ and/or $T$ is increased.  
We emphasize that our results can be expected to hold even when there is a small band gap, 
provided that this gap is smaller than either $\kb T$ or $\Ef$.

One can estimate quantitatively the expected thermopower and figure of merit for Pb$_{1-x}$Sn$_x$Se under generic experimental conditions using Eq.\ (\ref{eq:alphaEQLDirac}).  Inserting the measured value of the Dirac velocity \cite{liang_evidence_2013} gives
\be 
\alpha_{xx} \approx \left( 0.4 \frac{\mu\textrm{V}}{\textrm{K}} \right) \times \frac{ (T \textrm{ [K]}) (B \textrm{ [T]})}{ n \textrm{ [}10^{17} \textrm{ cm}^{-3} \textrm{]}}.
\nonumber
\ee
So, for example, a Pb$_{1-x}$Sn$_x$Se crystal with a doping concentration $n = 10^{17}$\,cm$^{-3}$ at temperature $T = 300$\,K and subjected to a magnetic field $B = 30$\,T can be expected to produce a thermopower $\alpha_{xx} \approx 3600$\,$\mu$V/K.  At such low doping, the Wiedemann-Franz law is strongly violated due to a phonon contribution to the thermal conductivity that is much larger than the electron contribution, and $\kappa_{xx}$ is of order $3\,\textrm{W/(m K)}$.  \cite{shulumba_intrinsic_2017}  The value of $\rho_{xx}$ can be estimated from the measurements of Ref.\ \cite{liang_evidence_2013}, which show a $B$-independent mobility $\mu_e$ that reaches $\approx 10^{5}$\,cm$^2 \,\textrm{V}^{-1} \textrm{s}^{-1}$ at zero temperature and that declines as $\mu_e \approx (1.5 \times 10^6\, \textrm{cm}^2\, \textrm{V}^{-1} \textrm{s}^{-1})/(T \textrm{ [K]})$ at temperatures above $\approx 20$\,K. 
(This result for $\rho_{xx}$ is consistent with previous measurements \cite{dixon_influence_1969, dziawa_topological_2012}.)  
Inserting these measurements into Eq.\ (\ref{eq:ZBTphonons}), and using $\rho_{xx} = 1/(n e \mu_{e})$, gives a figure of merit
\be 
Z_BT \approx 1.3 \times 10^{-7} \times \frac{ (T \textrm{ [K]})^2 (B \textrm{ [T]})^2}{ n \textrm{ [}10^{17} \textrm{ cm}^{-3} \textrm{]}}.
\nonumber
\ee
So, for example, at $n = 10^{17}$\,cm$^{-3}$, $T = 300$\,K, and $B = 30$\,T, the figure of merit can apparently reach an unprecedented value $Z_BT \approx 10$.  Such experimental conditions are already achievable in the laboratory, so that our results suggest an immediate pathway for arriving at record-large figure of merit.  Indeed, the sample studied in Ref.\ \cite{liang_evidence_2013} has $n \approx 3.5 \times 10^{17}$\,cm$^{-3}$, so that at $B = 30$\,T and $T = 300$\,K this sample should already exhibit $Z_B T \approx 3$.  If the doping concentration can be reduced as low as $n = 3 \times 10^{15}$\,cm$^{-3}$ (as has been achieved, for example, in the Dirac semimetals ZrTe$_5$ \cite{li_chiral_2016, liu_zeeman_2016} and HfTe$_5$  \cite{wang_chiral_2016}), then one can expect the room-temperature figure of merit to be larger than unity already at $B > 1$\,T.  The corresponding power factor is also enormously enhanced by the magnetic field,
\be 
\pf \approx \left( 4 \times 10^{-3} \frac{\mu\textrm{W}}{\textrm{cm} \textrm{ K}^2} \right) \times \frac{ (T \textrm{ [K]}) (B \textrm{ [T]})^2}{ n \textrm{ [}10^{17} \textrm{ cm}^{-3} \textrm{]}},
\nonumber
\ee 
reaching $\pf \approx 1000 \ \mu \textrm{W}/(\textrm{cm\,K}^2)$ at $n = 10^{17}$\,cm$^{-3}$, $T = 300$\,K, and $B = 30$\,T.

Finally, it is interesting to notice that Eq.\ (\ref{eq:alphaEQLDirac}) implies a thermopower that is largest in materials with low Dirac velocity and high valley degeneracy.  In this sense there appears to be considerable overlap between the search for effective thermoelectrics and the search for novel correlated electronic states.

\

\acknowledgments

We are grateful to Jiawei Zhou, Gang Chen, and Itamar Kimchi for helpful discussions.
BS was supported as part of the MIT Center for Excitonics, an Energy Frontier Research Center funded by the U.S. Department of Energy, Office of Science, Basic Energy Sciences under Award no.\ DE-SC0001088. LF's research is supported as part of the Solid-State Solar-Thermal Energy Conversion Center (S3TEC), an Energy Frontier Research Center funded by the U.S. Department of Energy (DOE), Office of Science, Basic Energy Sciences (BES), under Award DE-SC0001299 / DE-FG02-09ER46577 (thermoelectricity), and DOE Office of Basic Energy Sciences, Division of Materials Sciences and Engineering under Award DE-SC0010526 (topological materials). 

\pagebreak

\appendix

\widetext

\section{Generalized expression for the thermoelectric figure of merit and power factor}
\label{app:ZBT}

The figure of merit and power factor for a thermoelectric material can be derived, in general, by considering the thermodynamic efficiency of a thermoelectric generator or refrigerator. \cite{heikes_thermoelectricity:_1961} In the typical treatment, it is assumed that all response coefficients, including the electrical conductivity tensor $\shat$, the thermoelectric tensor $\ahat$, and thermal conductivity tensor $\khat$, are diagonal.  Here we briefly present a generalized derivation of the figure of merit and power factor that allows for all of these tensors to have off-diagonal components, and we derive the relevant expression for the maximal thermodynamic efficiency and power output.

\subsection{Transport Equations}

\begin{figure}[htb]
\centering
\includegraphics[width=0.85 \textwidth]{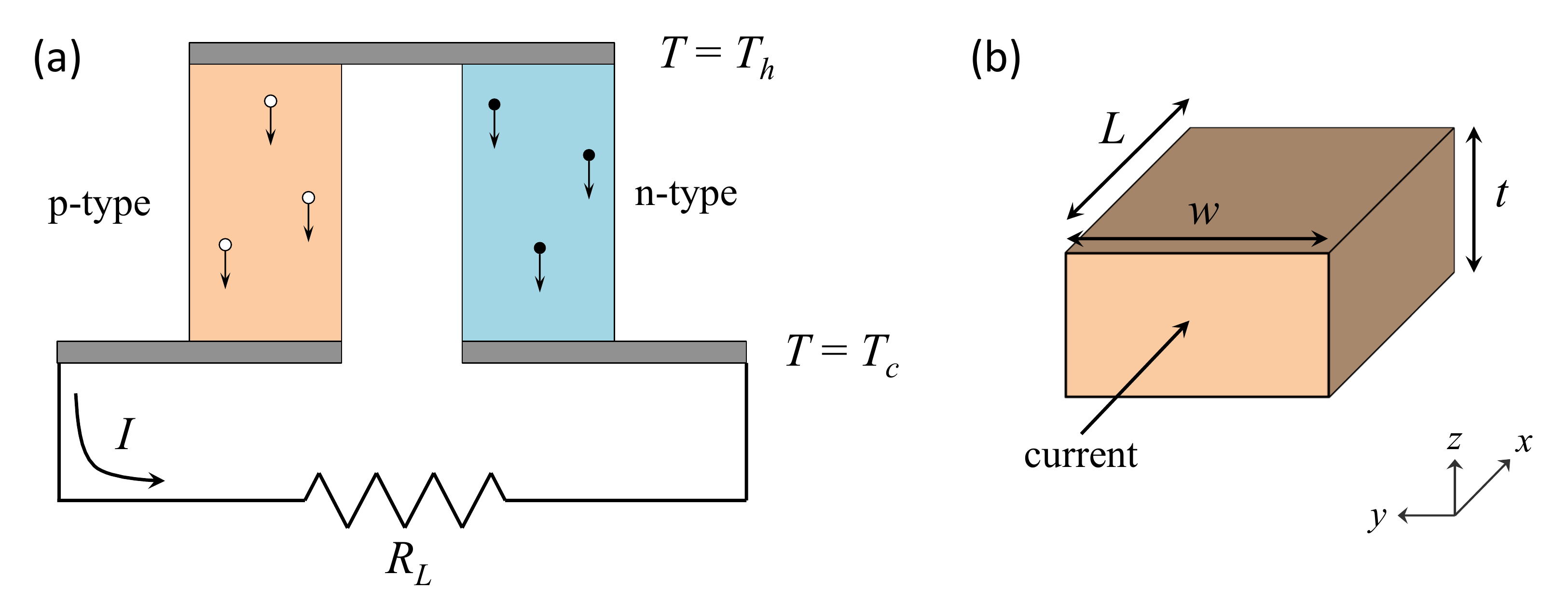}
\caption{(a) Typical setup of a thermoelectric generator module. The load resistance $R_L$ is tuned to produce optimal thermodynamic efficiency. (b) Dimensions and setup of a single leg of the module.}
\label{fig:device_schematic}
\end{figure}

Consider the typical setup for a thermoelectric generator module, shown schematically in Fig.\ \ref{fig:device_schematic}(a).  In this setup, an n-type and a p-type material are arranged to be in series with respect to electrical current and in parallel with respect to thermal current.  For simplicity, we assume that the two materials are identical except for the sign of the carrier doping, so that their response coefficients are identical up to the overall sign of the thermoelectric tensor and the off-diagonal components of the resistivity tensor.

Within linear response, the equations that dictate the flow of electric and heat current are
\begin{align}
\bE & = \rhat \bJ^e + \ahat \bold{\nabla} T \nonumber \\
\bJ^Q & = \Phat \bJ^e - \khat \bold{\nabla} T.
\label{eq:transport}
\end{align}
Here, $\bE$ is the electric field, $\bJ^e$ is the electrical current density, $\bJ^Q$ is the heat current density, $\ahat$ is the thermoelectric tensor, $T$ is the temperature, $\Phat$ is the Peltier tensor, and $\khat$ is the thermal conductivity tensor.  The Peltier tensor is related to the thermoelectric tensor via an Onsager reciprocal relation, $\Pi_{ij}(\bB) = -T \alpha_{ji}(-\bB)$, where $\bB$ is the magnetic field.  Since the matrix $\ahat$ is antisymmetric and its off-diagonal components must change sign under reversal of $\bB$, we generically have $\Phat = T \ahat$. Thus, we write the four response tensors as
\begin{align*}
 & & & &
 \rhat & =
  \begin{bmatrix}
   \rho_{xx} & \rho_{xy} \\
   -\rho_{xy} & \rho_{xx} \\
  \end{bmatrix}, 
	& \ahat &= 
	\begin{bmatrix}
	\alpha_{xx} & \alpha_{xy} \\
	-\alpha_{xy} & \alpha_{xx} \\
	\end{bmatrix}
& & & &
\\
& &	& & 
\Phat &=
  T \begin{bmatrix}
   \alpha_{xx} & \alpha_{xy} \\
   -\alpha_{xy} & \alpha_{xx} \\
  \end{bmatrix}, 
	& \khat &= 
	\begin{bmatrix}
	\kappa_{xx} & -\kappa_{xy} \\
	\kappa_{xy} & \kappa_{xx} \\
	\end{bmatrix} .
& &	& &
\end{align*}

We now focus on the electrical and thermal current through a single leg of the device.
We are interested in the situation where both the electrical and heat currents flow  uniformly in the $x$ direction [see Fig.\ \ref{fig:device_schematic}(b)], so that $\bJ^e = (I/w t) \hat{x}$ and $\bJ^Q = (Q/w t) \hat{x}$, where $I$ is the total electrical current and $Q$ is the total heat current.  Defining the resistance $R_{xx} = \rho_{xx} L/(w t)$, and then multiplying out the $x$-component of the first line of Eq.\ (\ref{eq:transport}) gives
\be
V = \alpha_{xx} (T_h - T_c) - I R_{xx} - \alpha_{xy} \Delta T_y \frac{L}{w}
\label{eq:V}
\ee 
for the voltage drop $V$ across the leg in the $x$ direction.  Here $\Delta T_y$ represents the difference in temperature in the transverse direction.  One can find the value of $\Delta T_y$ by examining the $y$-component of the second line of Eq.\ (\ref{eq:transport}), which gives
\be 
\Delta T_y = (T_h - T_c) \frac{\kappa_{xy}}{\kappa_{xx}} \frac{w}{L} - \frac{\alpha_{xy}}{\kappa_{xx} t} T I.
\label{eq:dTy}
\ee
Here, $T$ represents the average temperature $(T_h + T_c)/2$.  Finally, we can use the $x$-component of the second line in Eq.\ (\ref{eq:transport}) to define the total heat current $Q$ entering the leg from the hot junction, which is given by
\be 
\begin{split}
Q & = \alpha_{xx} T_h I - \kappa_{xx} \frac{w t}{L} (T_h - T_c) + \kappa_{xy} t \Delta T_y \\
& = T_h I \left( \alpha_{xx} - \alpha_{xy} \frac{\kappa_{xy}}{\kappa_{xx}} \right) + (T_h - T_c) \left( \kappa_{xx} + \frac{\kappa_{xy}^2}{\kappa_{xx}} \right) \frac{w t}{L}.
\end{split}
\label{eq:Q}
\ee
Here we have neglected the correction to the heat current associated with Joule heating within the sample, which is equivalent to considering only first-order terms in $(T_h - T_c)$. \cite{heikes_thermoelectricity:_1961}  

Since the two legs of the module are connected in series, the current through the sample is related to the voltage $V$ by $I = 2V/R_L$, where $R_L$ is the load resistance of the circuit.  Substituting Eq.\ (\ref{eq:dTy}) into Eq.\ (\ref{eq:V}), one can use this relation to solve for the current $I$, which gives
\be 
I = 2(T_h - T_c) \frac{\alpha_{xx} - \alpha_{xy} \frac{\kappa_{xy}}{\kappa_{xx}}}{2\left(\rho_{xx} - \frac{\alpha_{xy}^2}{\kappa_{xx}} T \right) \frac{L}{w t} + R_L}.
\label{eq:I}
\ee 

It is now convenient to define the following renormalized variables:
\begin{align}
\begin{split}
\ta & = \alpha_{xx} - \alpha_{xy} \frac{\kappa_{xy}}{\kappa_{xx}} \\
\tK &= \left( \kappa_{xx} + \frac{\kappa_{xy}^2}{\kappa_{xx}} \right) \frac{w t}{L} \\
\tR &= \left(\rho_{xx} - \frac{\alpha_{xy}^2}{\kappa_{xx}} T \right) \frac{L}{w t},
\end{split}
\end{align}
so that Eqs.\ (\ref{eq:Q}) and (\ref{eq:I}) become
\begin{align}
Q & = \ta T_h I + (T_h - T_c) \tK \\
I & = \frac{2 (T_h - T_c) \ta}{2 \tR + R_L} .
\end{align}
In this language, the expressions for the heat and electrical current through the leg have the same form as in the usual case, with only a renormalization to the transport coefficients arising from off-diagonal response. Indeed, in the absence of any off-diagonal coefficients, $\tK$ and $\tR$ are precisely the thermal conductance and the electrical resistance.

\subsection{Figure of Merit}

For optimal performance of the module, the load resistance $R_L$ should be tuned to maximize the thermodynamic efficiency
\be 
\eta = \frac{I^2 R_L}{2 Q}.
\ee 
The numerator of this equation represents the electrical work extracted from the module and the denominator is the total heat flowing into both legs from the hot junction.

Following the usual optimization of the load resistance to provide maximal $\eta$ (setting $d\eta/dR_L = 0$ and solving for $R_L$), we arrive at an optimal load resistance 
\be 
R_L^{\text{(opt)}} = 2 \tR \sqrt{1 + \tZ T},
\ee 
where $\tZ T$ is the effective figure of merit:
\be 
\tZ T = \frac{\ta^2 T}{\tK \tR} = \frac{ \alpha_{xx}^2 \left( 1 - \frac{\alpha_{xy}}{\alpha_{xx}} \frac{\kappa_{xy}}{\kappa_{xx}} \right)^2}
{\kappa_{xx} \rho_{xx} \left( 1 + \frac{\kappa_{xy}^2}{\kappa_{xx}^2} \right) \left( 1 - \frac{\alpha_{xy}^2 T}{\kappa_{xx} \rho_{xx}} \right) }.
\ee
This expression is equivalent to Eq.\ (3) of the main text.  

As in the usual case, the optimal module efficiency is given by
\be 
\eta^{\text{(opt)}} = \eta\left( R_L = R_L^{\text{(opt)}} \right) =  \frac{T_h - T_c}{T_h} \frac{\sqrt{1 + \tZ T} + 1}{\sqrt{1 + \tZ T} - 1}.
\ee 
As expected, when the figure of merit diverges, $\tZ T \rightarrow \infty$, the efficiency approaches the Carnot limit, $(T_h - T_c)/T_h$.

\subsection{Power Factor}

The corresponding expression for the power factor $\pf$ can be derived by considering the maximal electrical power that can be extracted for a given temperature difference $T_h - T_c$.  In particular, setting $d(I^2 R_L)/dR_L = 0$ and solving for $R_L$ gives the usual load matching condition, $R_L = 2 \tR$.  The corresponding electrical power
\be 
I^2 R_L = \frac{1}{2} (T_h - T_c)^2 \frac{\ta^2}{\tR},
\ee 
from which we can define the \emph{power factor}
\be 
\pf = \frac{\ta^2}{\tR \frac{wt}{L}}  = \frac{\left(\alpha_{xx} - \alpha_{xy} \frac{\kappa_{xy}}{\kappa_{xx}} \right)^2}{\rho_{xx} - \frac{\alpha_{xy}^2 T}{\kappa_{xx}} }.
\ee
One can think of $\pf$ as the maximal amount of useful electrical power per unit area that can be extracted for a given squared temperature difference, $(T_h - T_c)^2$.

\section{General expression for the thermopower of heavily-doped semiconductors}
\label{app:semiconductor}

In a quantizing magnetic field, the orbital degeneracy of each Landau level is given by the number of flux quanta passing through the system. The number of carriers (say, electrons) per flux quantum per unit length in the field direction is given by $n_1 = 2 \pi n \lb^2 n$.  The energy $\e$ of an electron eigenstate, relative to the band edge, is determined by the Landau level index $\ell$, by the momentum $\hbar k$ in the field direction, and by the spin $\zeta$:
\be
\e_{k\ell \zeta} = \frac{\hbar^2 k^2}{2 m} + \hbar \omega_c \left( \ell + \frac{1}{2} \right) - g \mub B \zeta.
\ee
Here $g$ denotes the electron g-factor, $\mub$ is the Bohr magneton, and $\zeta = \pm 1/2$.  For a given electron concentration $n$ the chemical potential $\mu$ of electrons is fixed by the relation
\be
N_v \int_{-\infty}^{\infty} \frac{dk}{2\pi} \sum_{\ell = 0}^\infty \sum_{\zeta = \pm \frac{1}{2}} f(\e_{k \ell \zeta}) = n_1,
\label{eq:mun1}
\ee
where $f(\e) = \{1 + \exp[(\e - \mu)/\kb T]\}^{-1}$ is the Fermi function and $N_v$ denotes the degeneracy of each spin/momentum state (the number of valleys).  Evaluating the integral over $k$ and the sum over $\zeta$ gives a self-consistency relation for the chemical potential:
\be
- \sum_{\ell = 0}^\infty \left\{
\textrm{Li}_{1/2} \left[ - \exp \left( \frac{(\mu - \frac{1}{2} \hbar \omega_c) - \frac{1}{2}g \mub B  - \hbar \omega_c \ell)}{\kb T} \right) \right]
 +
\textrm{Li}_{1/2} \left[ - \exp \left( \frac{(\mu - \frac{1}{2} \hbar \omega_c) + \frac{1}{2} g \mub B - \hbar \omega_c \ell}{\kb T} \right)  \right]
\right\}
=
\sqrt{\frac{2 \pi \hbar^2 n_1^2}{m \kb T N_v^2} }.
\label{eq:musemi}
\ee
Here $\textrm{Li}_{1/2}(x) = \sum_{j = 1}^\infty (x^j/\sqrt{j})$ is a polylogarithm function.  To produce the calculation shown in Fig.\ 2 we first solve Eq.\ (\ref{eq:musemi}) numerically for $\mu$ to determine the chemical potential at arbitrary values of $B$ and $T$.

The entropy per unit length per flux quantum is given by
\be
s_1 = - \kb N_v \int_{-\infty}^{\infty} \frac{dk}{2 \pi} \sum_{\ell = 0}^{\infty} \sum_{\zeta = \pm \frac{1}{2}} \left[f \ln f + (1 - f) \ln(1-f) \right].
\label{eq:s1}
\ee
where $f$ denotes $f(\e_{k \ell \zeta})$.  Dividing $s_1$ by $2 \pi \lb^2$ gives the total entropy per unit volume, and one can then use Eq.\ (2) of the main text to arrive at the following expression for the Seebeck coefficient $\alpha_{xx}$:
\be
\alpha_{xx} = \frac{N_v}{2\sqrt{2} \pi^2} \frac{\kb}{e} \sqrt{ \frac{m \kb T}{\hbar^2 n^2 \lb^4} } \sum_{\zeta = \pm \frac{1}{2} } \sum_{\ell = 0}^{\infty} \int_{-\infty}^{\infty}  \left[ \ln( e^y + 1) - \frac{y e^y}{e^y + 1} \right] dx.
\label{eq:alphasemi}
\ee
Here, $y = x^2 + [\hbar \omega_c (\ell + 1/2) + g \mub B \zeta - \mu]/(\kb T)$ is the electron energy in units of $\kb T$.  In Fig.\ 2 of the main text we show a numeric evaluation of Eq.\ (\ref{eq:alphasemi}) for the case of $\kb T = 0.02 \Ef^{(0)}$ as a function of magnetic field.

\section{General expression for the thermopower of Dirac/Weyl semimetals}
\label{app:Dirac}

As in the semiconductor case, we can calculate the thermopower for Dirac/Weyl semimetals by first determining the chemical potential at a given $B$ and $T$ and then calculating the entropy; only the form of the dispersion relation $\e_{k \ell}$ is different relative to the semiconductor case.

In particular, in Dirac/Weyl semimetals the single-particle energy levels in a magnetic field are given by\cite{jeon_landau_2014}
\be
\e_{k\ell} = \textrm{sign}(\ell) \times \sqrt{\frac{2 \hbar^2 v^2}{\lb^2} |\ell| + \hbar^2 v^2 k^2}.
\ee
This expression assumes that the energy scale for coupling of the field to the electron spin, $\sim \mub B$, is much smaller than the Landau level spacing $\sim \hbar v/\lb$. Unlike the usual case of a linear dispersion, for gapless Dirac materials the Landau level index $\ell$ can take any integer value $\ell = 0, \pm 1, \pm 2, $ etc.  The $\ell = 0$ level comprises one positive-dispersing branch with $\e = \hbar v k$ and one negative-dispersing branch with $\e = -\hbar v k$.

As in the semiconductor case, the chemical potential $\mu$ is fixed by the relation
\be
2 N_v \int_{-\infty}^{\infty} \frac{dk}{2\pi} \sum_{\ell = 0}^\infty  f(\e_{k \ell}) = 2 \pi n \lb^2.
\label{eq:mun1Dirac}
\ee
The $\ell = 0$ term in the sum is understood as taking one integral over the positively-dispersing branch $\e_k^{+} = \hbar v k$ and one integral over the negative-dispersing branch $\e_k^{-} = -\hbar v k$.  In other words, one can effectively replace the $\ell = 0$ term with $(1/2)[f(\e_k^{+}) + f(\e_k^{-})]$.
Equation (\ref{eq:mun1Dirac}) can be solved numerically for generic $B$ and $T$.

Once the chemical potential $\mu$ is known, the Seebeck coefficient $\alpha_{xx}$ can be determined by calculating the total electronic entropy and dividing by the net charge.  This procedure gives
\be
\alpha_{xx} = -\frac{\kb}{e} \frac{N_v}{\pi n \lb^2} \int_{-\infty}^{\infty} \frac{dk}{2\pi} \sum_{\ell = -\infty}^{\infty} \left[f \ln f + (1 - f) \ln (1 - f) \right].
\ee
As with Eq.\ (\ref{eq:mun1Dirac}), the $\ell = 0$ term of the sum can be interpreted as $(1/2)[f(\e_k^{+}) + f(\e_k^{-})]$.

\bibliography{thermopower}

\begin{thebibliography}{32}%
\makeatletter
\providecommand \@ifxundefined [1]{%
 \@ifx{#1\undefined}
}%
\providecommand \@ifnum [1]{%
 \ifnum #1\expandafter \@firstoftwo
 \else \expandafter \@secondoftwo
 \fi
}%
\providecommand \@ifx [1]{%
 \ifx #1\expandafter \@firstoftwo
 \else \expandafter \@secondoftwo
 \fi
}%
\providecommand \natexlab [1]{#1}%
\providecommand \enquote  [1]{``#1''}%
\providecommand \bibnamefont  [1]{#1}%
\providecommand \bibfnamefont [1]{#1}%
\providecommand \citenamefont [1]{#1}%
\providecommand \href@noop [0]{\@secondoftwo}%
\providecommand \href [0]{\begingroup \@sanitize@url \@href}%
\providecommand \@href[1]{\@@startlink{#1}\@@href}%
\providecommand \@@href[1]{\endgroup#1\@@endlink}%
\providecommand \@sanitize@url [0]{\catcode `\\12\catcode `\$12\catcode
  `\&12\catcode `\#12\catcode `\^12\catcode `\_12\catcode `\%12\relax}%
\providecommand \@@startlink[1]{}%
\providecommand \@@endlink[0]{}%
\providecommand \url  [0]{\begingroup\@sanitize@url \@url }%
\providecommand \@url [1]{\endgroup\@href {#1}{\urlprefix }}%
\providecommand \urlprefix  [0]{URL }%
\providecommand \Eprint [0]{\href }%
\providecommand \doibase [0]{http://dx.doi.org/}%
\providecommand \selectlanguage [0]{\@gobble}%
\providecommand \bibinfo  [0]{\@secondoftwo}%
\providecommand \bibfield  [0]{\@secondoftwo}%
\providecommand \translation [1]{[#1]}%
\providecommand \BibitemOpen [0]{}%
\providecommand \bibitemStop [0]{}%
\providecommand \bibitemNoStop [0]{.\EOS\space}%
\providecommand \EOS [0]{\spacefactor3000\relax}%
\providecommand \BibitemShut  [1]{\csname bibitem#1\endcsname}%
\let\auto@bib@innerbib\@empty
\bibitem [{\citenamefont {Ioffe}(1957)}]{ioffe_semiconductor_1957}%
  \BibitemOpen
  \bibfield  {author} {\bibinfo {author} {\bibfnamefont {A.~F.}\ \bibnamefont
  {Ioffe}},\ }\href@noop {} {\emph {\bibinfo {title} {Semiconductor
  {Thermoelements} and {Thermo}-electric {Cooling}}}}\ (\bibinfo  {publisher}
  {Infosearch},\ \bibinfo {address} {London},\ \bibinfo {year}
  {1957})\BibitemShut {NoStop}%
\bibitem [{\citenamefont {Dresselhaus}\ \emph {et~al.}(2007)\citenamefont
  {Dresselhaus}, \citenamefont {Chen}, \citenamefont {Tang}, \citenamefont
  {Yang}, \citenamefont {Lee}, \citenamefont {Wang}, \citenamefont {Ren},
  \citenamefont {Fleurial},\ and\ \citenamefont
  {Gogna}}]{dresselhaus_new_2007}%
  \BibitemOpen
  \bibfield  {author} {\bibinfo {author} {\bibfnamefont {M.~S.}\ \bibnamefont
  {Dresselhaus}}, \bibinfo {author} {\bibfnamefont {G.}~\bibnamefont {Chen}},
  \bibinfo {author} {\bibfnamefont {M.~Y.}\ \bibnamefont {Tang}}, \bibinfo
  {author} {\bibfnamefont {R.~G.}\ \bibnamefont {Yang}}, \bibinfo {author}
  {\bibfnamefont {H.}~\bibnamefont {Lee}}, \bibinfo {author} {\bibfnamefont
  {D.~Z.}\ \bibnamefont {Wang}}, \bibinfo {author} {\bibfnamefont {Z.~F.}\
  \bibnamefont {Ren}}, \bibinfo {author} {\bibfnamefont {J.-P.}\ \bibnamefont
  {Fleurial}}, \ and\ \bibinfo {author} {\bibfnamefont {P.}~\bibnamefont
  {Gogna}},\ }\bibfield  {title} {\enquote {\bibinfo {title} {New {Directions}
  for {Low}-{Dimensional} {Thermoelectric} {Materials}},}\ }\href {\doibase
  10.1002/adma.200600527} {\bibfield  {journal} {\bibinfo  {journal} {Advanced
  Materials}\ }\textbf {\bibinfo {volume} {19}},\ \bibinfo {pages} {1043--1053}
  (\bibinfo {year} {2007})}\BibitemShut {NoStop}%
\bibitem [{\citenamefont {Shakouri}(2011)}]{shakouri_recent_2011}%
  \BibitemOpen
  \bibfield  {author} {\bibinfo {author} {\bibfnamefont {Ali}\ \bibnamefont
  {Shakouri}},\ }\bibfield  {title} {\enquote {\bibinfo {title} {Recent
  {Developments} in {Semiconductor} {Thermoelectric} {Physics} and
  {Materials}},}\ }\href {\doibase 10.1146/annurev-matsci-062910-100445}
  {\bibfield  {journal} {\bibinfo  {journal} {Annual Review of Materials
  Research}\ }\textbf {\bibinfo {volume} {41}},\ \bibinfo {pages} {399--431}
  (\bibinfo {year} {2011})}\BibitemShut {NoStop}%
\bibitem [{\citenamefont {Chen}\ and\ \citenamefont
  {Shklovskii}(2013)}]{Chen_anomalously_2013}%
  \BibitemOpen
  \bibfield  {author} {\bibinfo {author} {\bibfnamefont {Tianran}\ \bibnamefont
  {Chen}}\ and\ \bibinfo {author} {\bibfnamefont {B.~I.}\ \bibnamefont
  {Shklovskii}},\ }\bibfield  {title} {\enquote {\bibinfo {title} {Anomalously
  small resistivity and thermopower of strongly compensated semiconductors and
  topological insulators},}\ }\href {\doibase 10.1103/PhysRevB.87.165119}
  {\bibfield  {journal} {\bibinfo  {journal} {Phys. Rev. B}\ }\textbf {\bibinfo
  {volume} {87}},\ \bibinfo {pages} {165119} (\bibinfo {year}
  {2013})}\BibitemShut {NoStop}%
\bibitem [{\citenamefont {Mahan}(1989)}]{Mahan_figure_1989}%
  \BibitemOpen
  \bibfield  {author} {\bibinfo {author} {\bibfnamefont {G.~D.}\ \bibnamefont
  {Mahan}},\ }\bibfield  {title} {\enquote {\bibinfo {title} {Figure of merit
  for thermoelectrics},}\ }\href {\doibase 10.1063/1.342976} {\bibfield
  {journal} {\bibinfo  {journal} {Journal of Applied Physics}\ }\textbf
  {\bibinfo {volume} {65}},\ \bibinfo {pages} {1578} (\bibinfo {year}
  {1989})}\BibitemShut {NoStop}%
\bibitem [{\citenamefont {Obraztsov}(1965)}]{obraztsov_thermal_1965}%
  \BibitemOpen
  \bibfield  {author} {\bibinfo {author} {\bibfnamefont {Yu.~N.}\ \bibnamefont
  {Obraztsov}},\ }\bibfield  {title} {\enquote {\bibinfo {title} {The thermal
  {EMF} of semiconductors in a quantizing magnetic field},}\ }\href@noop {}
  {\bibfield  {journal} {\bibinfo  {journal} {Sov. Phys. - Solid State}\
  }\textbf {\bibinfo {volume} {7}},\ \bibinfo {pages} {455} (\bibinfo {year}
  {1965})}\BibitemShut {NoStop}%
\bibitem [{\citenamefont {Tsendin}\ and\ \citenamefont
  {Efros}(1966)}]{tsendin_theory_1966}%
  \BibitemOpen
  \bibfield  {author} {\bibinfo {author} {\bibfnamefont {K.~D.}\ \bibnamefont
  {Tsendin}}\ and\ \bibinfo {author} {\bibfnamefont {A.~L.}\ \bibnamefont
  {Efros}},\ }\bibfield  {title} {\enquote {\bibinfo {title} {Theory of thermal
  {EMF} in a quantizing magnetic field in the {Kane} model},}\ }\href@noop {}
  {\bibfield  {journal} {\bibinfo  {journal} {Sov. Phys. - Solid State}\
  }\textbf {\bibinfo {volume} {8}},\ \bibinfo {pages} {306} (\bibinfo {year}
  {1966})}\BibitemShut {NoStop}%
\bibitem [{\citenamefont {Jay-Gerin}(1974)}]{jay-gerin_thermoelectric_1974}%
  \BibitemOpen
  \bibfield  {author} {\bibinfo {author} {\bibfnamefont {J.~P.}\ \bibnamefont
  {Jay-Gerin}},\ }\bibfield  {title} {\enquote {\bibinfo {title}
  {Thermoelectric power of semiconductors in the extreme quantum limit. {I}.
  {The} ``electron-diffusion'' contribution.}}\ }\href {\doibase
  10.1016/0022-3697(74)90015-8} {\bibfield  {journal} {\bibinfo  {journal}
  {Journal of Physics and Chemistry of Solids}\ }\textbf {\bibinfo {volume}
  {35}},\ \bibinfo {pages} {81--87} (\bibinfo {year} {1974})}\BibitemShut
  {NoStop}%
\bibitem [{\citenamefont {Abrikosov}(1988)}]{abrikosov_fundamentals_1988}%
  \BibitemOpen
  \bibfield  {author} {\bibinfo {author} {\bibfnamefont {Alexei~A.}\
  \bibnamefont {Abrikosov}},\ }\href@noop {} {\emph {\bibinfo {title}
  {Fundamentals of the {Theory} of {Metals}}}}\ (\bibinfo  {publisher}
  {Elsevier},\ \bibinfo {address} {New York},\ \bibinfo {year}
  {1988})\BibitemShut {NoStop}%
\bibitem [{\citenamefont {Bergman}\ and\ \citenamefont
  {Oganesyan}(2010)}]{bergman_theory_2010}%
  \BibitemOpen
  \bibfield  {author} {\bibinfo {author} {\bibfnamefont {Doron~L.}\
  \bibnamefont {Bergman}}\ and\ \bibinfo {author} {\bibfnamefont {Vadim}\
  \bibnamefont {Oganesyan}},\ }\bibfield  {title} {\enquote {\bibinfo {title}
  {Theory of {Dissipationless} {Nernst} {Effects}},}\ }\href {\doibase
  10.1103/PhysRevLett.104.066601} {\bibfield  {journal} {\bibinfo  {journal}
  {Physical Review Letters}\ }\textbf {\bibinfo {volume} {104}},\ \bibinfo
  {pages} {066601} (\bibinfo {year} {2010})}\BibitemShut {NoStop}%
\bibitem [{\citenamefont {Ziman}(1972)}]{ziman_principles_1972}%
  \BibitemOpen
  \bibfield  {author} {\bibinfo {author} {\bibfnamefont {J.~M.}\ \bibnamefont
  {Ziman}},\ }\href@noop {} {\emph {\bibinfo {title} {Principles of the
  {Theory} of {Solids}}}}\ (\bibinfo  {publisher} {Cambridge University
  Press},\ \bibinfo {address} {New York},\ \bibinfo {year} {1972})\BibitemShut
  {NoStop}%
\bibitem [{\citenamefont {Jay-Gerin}(1975)}]{jay-gerin_thermoelectric_1975}%
  \BibitemOpen
  \bibfield  {author} {\bibinfo {author} {\bibfnamefont {J.~P.}\ \bibnamefont
  {Jay-Gerin}},\ }\bibfield  {title} {\enquote {\bibinfo {title}
  {Thermoelectric power of semiconductors in the extreme quantum limit. {II}.
  {The} ``phonon-drag" contribution},}\ }\href {\doibase
  10.1103/PhysRevB.12.1418} {\bibfield  {journal} {\bibinfo  {journal}
  {Physical Review B}\ }\textbf {\bibinfo {volume} {12}},\ \bibinfo {pages}
  {1418--1431} (\bibinfo {year} {1975})}\BibitemShut {NoStop}%
\bibitem [{\citenamefont {Pepper}(1979)}]{pepper_metal-insulator_1979}%
  \BibitemOpen
  \bibfield  {author} {\bibinfo {author} {\bibfnamefont {M.}~\bibnamefont
  {Pepper}},\ }\bibfield  {title} {\enquote {\bibinfo {title} {Metal-insulator
  transitions induced by a magnetic field},}\ }\href {\doibase
  10.1016/0022-3093(79)90071-1} {\bibfield  {journal} {\bibinfo  {journal}
  {Journal of Non-Crystalline Solids}\ }\textbf {\bibinfo {volume} {32}},\
  \bibinfo {pages} {161--185} (\bibinfo {year} {1979})}\BibitemShut {NoStop}%
\bibitem [{\citenamefont {Arora}\ and\ \citenamefont
  {Al-Missari}(1979)}]{arora_thermoelectric_1979}%
  \BibitemOpen
  \bibfield  {author} {\bibinfo {author} {\bibfnamefont {Vija~K.}\ \bibnamefont
  {Arora}}\ and\ \bibinfo {author} {\bibfnamefont {Mahommad~A.}\ \bibnamefont
  {Al-Missari}},\ }\bibfield  {title} {\enquote {\bibinfo {title}
  {Thermoelectric power in high magnetic fields},}\ }\href {\doibase
  10.1016/0304-8853(79)90239-7} {\bibfield  {journal} {\bibinfo  {journal}
  {Journal of Magnetism and Magnetic Materials}\ }\textbf {\bibinfo {volume}
  {11}},\ \bibinfo {pages} {80--83} (\bibinfo {year} {1979})}\BibitemShut
  {NoStop}%
\bibitem [{\citenamefont {Dingle}(1955)}]{dingle_scattering_1955}%
  \BibitemOpen
  \bibfield  {author} {\bibinfo {author} {\bibfnamefont {R.B.}\ \bibnamefont
  {Dingle}},\ }\bibfield  {title} {\enquote {\bibinfo {title} {Scattering of
  electrons and holes by charged donors and acceptors in semiconductors},}\
  }\href {\doibase 10.1080/14786440808561235} {\bibfield  {journal} {\bibinfo
  {journal} {Philosophical Magazine}\ }\textbf {\bibinfo {volume} {46}},\
  \bibinfo {pages} {831--840} (\bibinfo {year} {1955})}\BibitemShut {NoStop}%
\bibitem [{\citenamefont {Herring}(1955)}]{herring_transport_1955}%
  \BibitemOpen
  \bibfield  {author} {\bibinfo {author} {\bibfnamefont {Conyers}\ \bibnamefont
  {Herring}},\ }\bibfield  {title} {\enquote {\bibinfo {title} {Transport
  {Properties} of a {Many}‐{Valley} {Semiconductor}},}\ }\href {\doibase
  10.1002/j.1538-7305.1955.tb01472.x} {\bibfield  {journal} {\bibinfo
  {journal} {Bell System Technical Journal}\ }\textbf {\bibinfo {volume}
  {34}},\ \bibinfo {pages} {237--290} (\bibinfo {year} {1955})}\BibitemShut
  {NoStop}%
\bibitem [{\citenamefont {Skinner}(2014)}]{skinner_coulomb_2014}%
  \BibitemOpen
  \bibfield  {author} {\bibinfo {author} {\bibfnamefont {Brian}\ \bibnamefont
  {Skinner}},\ }\bibfield  {title} {\enquote {\bibinfo {title} {Coulomb
  disorder in three-dimensional {Dirac} systems},}\ }\href {\doibase
  10.1103/PhysRevB.90.060202} {\bibfield  {journal} {\bibinfo  {journal}
  {Physical Review B}\ }\textbf {\bibinfo {volume} {90}},\ \bibinfo {pages}
  {060202} (\bibinfo {year} {2014})}\BibitemShut {NoStop}%
\bibitem [{\citenamefont {Gurevich}(1995)}]{gurevich_nature_1995}%
  \BibitemOpen
  \bibfield  {author} {\bibinfo {author} {\bibfnamefont {Yu.~G.}\ \bibnamefont
  {Gurevich}},\ }\bibfield  {title} {\enquote {\bibinfo {title} {Nature of the
  thermopower in bipolar semiconductors},}\ }\href {\doibase
  10.1103/PhysRevB.51.6999} {\bibfield  {journal} {\bibinfo  {journal}
  {Physical Review B}\ }\textbf {\bibinfo {volume} {51}},\ \bibinfo {pages}
  {6999--7004} (\bibinfo {year} {1995})}\BibitemShut {NoStop}%
\bibitem [{\citenamefont {Ashcroft}\ and\ \citenamefont
  {Mermin}(1976)}]{ashcroft_solid_1976}%
  \BibitemOpen
  \bibfield  {author} {\bibinfo {author} {\bibfnamefont {N.~W.}\ \bibnamefont
  {Ashcroft}}\ and\ \bibinfo {author} {\bibfnamefont {N.~D.}\ \bibnamefont
  {Mermin}},\ }\href@noop {} {\emph {\bibinfo {title} {Solid {State}
  {Physics}}}}\ (\bibinfo  {publisher} {Holt, Rinehart and Winston},\ \bibinfo
  {address} {New York},\ \bibinfo {year} {1976})\BibitemShut {NoStop}%
\bibitem [{\citenamefont {Rosenbaum}\ \emph {et~al.}(1985)\citenamefont
  {Rosenbaum}, \citenamefont {Field}, \citenamefont {Nelson},\ and\
  \citenamefont {Littlewood}}]{rosenbaum_magnetic-field-induced_1985}%
  \BibitemOpen
  \bibfield  {author} {\bibinfo {author} {\bibfnamefont {T.~F.}\ \bibnamefont
  {Rosenbaum}}, \bibinfo {author} {\bibfnamefont {Stuart~B.}\ \bibnamefont
  {Field}}, \bibinfo {author} {\bibfnamefont {D.~A.}\ \bibnamefont {Nelson}}, \
  and\ \bibinfo {author} {\bibfnamefont {P.~B.}\ \bibnamefont {Littlewood}},\
  }\bibfield  {title} {\enquote {\bibinfo {title} {Magnetic-{Field}-{Induced}
  {Localization} {Transition} in {HgCdTe}},}\ }\href {\doibase
  10.1103/PhysRevLett.54.241} {\bibfield  {journal} {\bibinfo  {journal}
  {Physical Review Letters}\ }\textbf {\bibinfo {volume} {54}},\ \bibinfo
  {pages} {241--244} (\bibinfo {year} {1985})}\BibitemShut {NoStop}%
\bibitem [{\citenamefont {Shayegan}\ \emph {et~al.}(1988)\citenamefont
  {Shayegan}, \citenamefont {Goldman},\ and\ \citenamefont
  {Drew}}]{shayegan_magnetic-field-induced_1988}%
  \BibitemOpen
  \bibfield  {author} {\bibinfo {author} {\bibfnamefont {M.}~\bibnamefont
  {Shayegan}}, \bibinfo {author} {\bibfnamefont {V.~J.}\ \bibnamefont
  {Goldman}}, \ and\ \bibinfo {author} {\bibfnamefont {H.~D.}\ \bibnamefont
  {Drew}},\ }\bibfield  {title} {\enquote {\bibinfo {title}
  {Magnetic-field-induced localization in narrow-gap semiconductors
  {Hg}$_{1-x}${Cd}$_{x}${Te} and {InSb}},}\ }\href {\doibase
  10.1103/PhysRevB.38.5585} {\bibfield  {journal} {\bibinfo  {journal}
  {Physical Review B}\ }\textbf {\bibinfo {volume} {38}},\ \bibinfo {pages}
  {5585--5602} (\bibinfo {year} {1988})}\BibitemShut {NoStop}%
\bibitem [{\citenamefont {Kozuka}\ \emph {et~al.}(2008)\citenamefont {Kozuka},
  \citenamefont {Susaki},\ and\ \citenamefont {Hwang}}]{kozuka_vanishing_2008}%
  \BibitemOpen
  \bibfield  {author} {\bibinfo {author} {\bibfnamefont {Y.}~\bibnamefont
  {Kozuka}}, \bibinfo {author} {\bibfnamefont {T.}~\bibnamefont {Susaki}}, \
  and\ \bibinfo {author} {\bibfnamefont {H.~Y.}\ \bibnamefont {Hwang}},\
  }\bibfield  {title} {\enquote {\bibinfo {title} {Vanishing {Hall}
  {Coefficient} in the {Extreme} {Quantum} {Limit} in {Photocarrier}-{Doped}
  {SrTiO}$_3$},}\ }\href {\doibase 10.1103/PhysRevLett.101.096601} {\bibfield
  {journal} {\bibinfo  {journal} {Physical Review Letters}\ }\textbf {\bibinfo
  {volume} {101}},\ \bibinfo {pages} {096601} (\bibinfo {year}
  {2008})}\BibitemShut {NoStop}%
\bibitem [{\citenamefont {Bhattacharya}\ \emph {et~al.}(2016)\citenamefont
  {Bhattacharya}, \citenamefont {Skinner}, \citenamefont {Khalsa},\ and\
  \citenamefont {Suslov}}]{bhattacharya_spatially_2016}%
  \BibitemOpen
  \bibfield  {author} {\bibinfo {author} {\bibfnamefont {Anand}\ \bibnamefont
  {Bhattacharya}}, \bibinfo {author} {\bibfnamefont {Brian}\ \bibnamefont
  {Skinner}}, \bibinfo {author} {\bibfnamefont {Guru}\ \bibnamefont {Khalsa}},
  \ and\ \bibinfo {author} {\bibfnamefont {Alexey~V.}\ \bibnamefont {Suslov}},\
  }\bibfield  {title} {\enquote {\bibinfo {title} {Spatially inhomogeneous
  electron state deep in the extreme quantum limit of strontium titanate},}\
  }\href {\doibase 10.1038/ncomms12974} {\bibfield  {journal} {\bibinfo
  {journal} {Nature Communications}\ }\textbf {\bibinfo {volume} {7}},\
  \bibinfo {pages} {12974} (\bibinfo {year} {2016})}\BibitemShut {NoStop}%
\bibitem [{\citenamefont {Liang}\ \emph {et~al.}(2013)\citenamefont {Liang},
  \citenamefont {Gibson}, \citenamefont {Xiong}, \citenamefont {Hirschberger},
  \citenamefont {Koduvayur}, \citenamefont {Cava},\ and\ \citenamefont
  {Ong}}]{liang_evidence_2013}%
  \BibitemOpen
  \bibfield  {author} {\bibinfo {author} {\bibfnamefont {Tian}\ \bibnamefont
  {Liang}}, \bibinfo {author} {\bibfnamefont {Quinn}\ \bibnamefont {Gibson}},
  \bibinfo {author} {\bibfnamefont {Jun}\ \bibnamefont {Xiong}}, \bibinfo
  {author} {\bibfnamefont {Max}\ \bibnamefont {Hirschberger}}, \bibinfo
  {author} {\bibfnamefont {Sunanda~P.}\ \bibnamefont {Koduvayur}}, \bibinfo
  {author} {\bibfnamefont {R.~J.}\ \bibnamefont {Cava}}, \ and\ \bibinfo
  {author} {\bibfnamefont {N.~P.}\ \bibnamefont {Ong}},\ }\bibfield  {title}
  {\enquote {\bibinfo {title} {Evidence for massive bulk {Dirac} fermions in
  {Pb}$_{1−x}${Sn}$_{x}${Se} from {Nernst} and thermopower experiments},}\
  }\href {\doibase 10.1038/ncomms3696} {\bibfield  {journal} {\bibinfo
  {journal} {Nature Communications}\ }\textbf {\bibinfo {volume} {4}},\
  \bibinfo {pages} {2696} (\bibinfo {year} {2013})}\BibitemShut {NoStop}%
\bibitem [{\citenamefont {Shulumba}\ \emph {et~al.}(2017)\citenamefont
  {Shulumba}, \citenamefont {Hellman},\ and\ \citenamefont
  {Minnich}}]{shulumba_intrinsic_2017}%
  \BibitemOpen
  \bibfield  {author} {\bibinfo {author} {\bibfnamefont {Nina}\ \bibnamefont
  {Shulumba}}, \bibinfo {author} {\bibfnamefont {Olle}\ \bibnamefont
  {Hellman}}, \ and\ \bibinfo {author} {\bibfnamefont {Austin~J.}\ \bibnamefont
  {Minnich}},\ }\bibfield  {title} {\enquote {\bibinfo {title} {Intrinsic
  localized mode and low thermal conductivity of {PbSe}},}\ }\href {\doibase
  10.1103/PhysRevB.95.014302} {\bibfield  {journal} {\bibinfo  {journal}
  {Physical Review B}\ }\textbf {\bibinfo {volume} {95}},\ \bibinfo {pages}
  {014302} (\bibinfo {year} {2017})}\BibitemShut {NoStop}%
\bibitem [{\citenamefont {Dixon}\ and\ \citenamefont
  {Hoff}(1969)}]{dixon_influence_1969}%
  \BibitemOpen
  \bibfield  {author} {\bibinfo {author} {\bibfnamefont {J.~R.}\ \bibnamefont
  {Dixon}}\ and\ \bibinfo {author} {\bibfnamefont {G.~F.}\ \bibnamefont
  {Hoff}},\ }\bibfield  {title} {\enquote {\bibinfo {title} {Influence of band
  inversion upon the electrical properties of {Pb}$_x${Sn}$_{1−x}${Se} in the
  low carrier concentration range},}\ }\href {\doibase
  10.1016/0038-1098(69)90283-X} {\bibfield  {journal} {\bibinfo  {journal}
  {Solid State Communications}\ }\textbf {\bibinfo {volume} {7}},\ \bibinfo
  {pages} {1777--1779} (\bibinfo {year} {1969})}\BibitemShut {NoStop}%
\bibitem [{\citenamefont {Dziawa}\ \emph {et~al.}(2012)\citenamefont {Dziawa},
  \citenamefont {Kowalski}, \citenamefont {Dybko}, \citenamefont {Buczko},
  \citenamefont {Szczerbakow}, \citenamefont {Szot}, \citenamefont
  {Łusakowska}, \citenamefont {Balasubramanian}, \citenamefont {Wojek},
  \citenamefont {Berntsen}, \citenamefont {Tjernberg},\ and\ \citenamefont
  {Story}}]{dziawa_topological_2012}%
  \BibitemOpen
  \bibfield  {author} {\bibinfo {author} {\bibfnamefont {P.}~\bibnamefont
  {Dziawa}}, \bibinfo {author} {\bibfnamefont {B.~J.}\ \bibnamefont
  {Kowalski}}, \bibinfo {author} {\bibfnamefont {K.}~\bibnamefont {Dybko}},
  \bibinfo {author} {\bibfnamefont {R.}~\bibnamefont {Buczko}}, \bibinfo
  {author} {\bibfnamefont {A.}~\bibnamefont {Szczerbakow}}, \bibinfo {author}
  {\bibfnamefont {M.}~\bibnamefont {Szot}}, \bibinfo {author} {\bibfnamefont
  {E.}~\bibnamefont {Łusakowska}}, \bibinfo {author} {\bibfnamefont
  {T.}~\bibnamefont {Balasubramanian}}, \bibinfo {author} {\bibfnamefont
  {B.~M.}\ \bibnamefont {Wojek}}, \bibinfo {author} {\bibfnamefont {M.~H.}\
  \bibnamefont {Berntsen}}, \bibinfo {author} {\bibfnamefont {O.}~\bibnamefont
  {Tjernberg}}, \ and\ \bibinfo {author} {\bibfnamefont {T.}~\bibnamefont
  {Story}},\ }\bibfield  {title} {\enquote {\bibinfo {title} {Topological
  crystalline insulator states in
  {Pb}$_{\textrm{1-\textit{x}}}${Sn}$_{\textrm{\textit{x}}}${Se}},}\ }\href
  {\doibase 10.1038/nmat3449} {\bibfield  {journal} {\bibinfo  {journal}
  {Nature Materials}\ }\textbf {\bibinfo {volume} {11}},\ \bibinfo {pages}
  {1023} (\bibinfo {year} {2012})}\BibitemShut {NoStop}%
\bibitem [{\citenamefont {Li}\ \emph {et~al.}(2016)\citenamefont {Li},
  \citenamefont {Kharzeev}, \citenamefont {Zhang}, \citenamefont {Huang},
  \citenamefont {Pletikosić}, \citenamefont {Fedorov}, \citenamefont {Zhong},
  \citenamefont {Schneeloch}, \citenamefont {Gu},\ and\ \citenamefont
  {Valla}}]{li_chiral_2016}%
  \BibitemOpen
  \bibfield  {author} {\bibinfo {author} {\bibfnamefont {Qiang}\ \bibnamefont
  {Li}}, \bibinfo {author} {\bibfnamefont {Dmitri~E.}\ \bibnamefont
  {Kharzeev}}, \bibinfo {author} {\bibfnamefont {Cheng}\ \bibnamefont {Zhang}},
  \bibinfo {author} {\bibfnamefont {Yuan}\ \bibnamefont {Huang}}, \bibinfo
  {author} {\bibfnamefont {I.}~\bibnamefont {Pletikosić}}, \bibinfo {author}
  {\bibfnamefont {A.~V.}\ \bibnamefont {Fedorov}}, \bibinfo {author}
  {\bibfnamefont {R.~D.}\ \bibnamefont {Zhong}}, \bibinfo {author}
  {\bibfnamefont {J.~A.}\ \bibnamefont {Schneeloch}}, \bibinfo {author}
  {\bibfnamefont {G.~D.}\ \bibnamefont {Gu}}, \ and\ \bibinfo {author}
  {\bibfnamefont {T.}~\bibnamefont {Valla}},\ }\bibfield  {title} {\enquote
  {\bibinfo {title} {Chiral magnetic effect in {ZrTe}5},}\ }\href {\doibase
  10.1038/nphys3648} {\bibfield  {journal} {\bibinfo  {journal} {Nature
  Physics}\ }\textbf {\bibinfo {volume} {12}},\ \bibinfo {pages} {550--554}
  (\bibinfo {year} {2016})}\BibitemShut {NoStop}%
\bibitem [{\citenamefont {Liu}\ \emph {et~al.}(2016)\citenamefont {Liu},
  \citenamefont {Yuan}, \citenamefont {Zhang}, \citenamefont {Jin},
  \citenamefont {Narayan}, \citenamefont {Luo}, \citenamefont {Chen},
  \citenamefont {Yang}, \citenamefont {Zou}, \citenamefont {Wu}, \citenamefont
  {Sanvito}, \citenamefont {Xia}, \citenamefont {Li}, \citenamefont {Wang},\
  and\ \citenamefont {Xiu}}]{liu_zeeman_2016}%
  \BibitemOpen
  \bibfield  {author} {\bibinfo {author} {\bibfnamefont {Yanwen}\ \bibnamefont
  {Liu}}, \bibinfo {author} {\bibfnamefont {Xiang}\ \bibnamefont {Yuan}},
  \bibinfo {author} {\bibfnamefont {Cheng}\ \bibnamefont {Zhang}}, \bibinfo
  {author} {\bibfnamefont {Zhao}\ \bibnamefont {Jin}}, \bibinfo {author}
  {\bibfnamefont {Awadhesh}\ \bibnamefont {Narayan}}, \bibinfo {author}
  {\bibfnamefont {Chen}\ \bibnamefont {Luo}}, \bibinfo {author} {\bibfnamefont
  {Zhigang}\ \bibnamefont {Chen}}, \bibinfo {author} {\bibfnamefont {Lei}\
  \bibnamefont {Yang}}, \bibinfo {author} {\bibfnamefont {Jin}\ \bibnamefont
  {Zou}}, \bibinfo {author} {\bibfnamefont {Xing}\ \bibnamefont {Wu}}, \bibinfo
  {author} {\bibfnamefont {Stefano}\ \bibnamefont {Sanvito}}, \bibinfo {author}
  {\bibfnamefont {Zhengcai}\ \bibnamefont {Xia}}, \bibinfo {author}
  {\bibfnamefont {Liang}\ \bibnamefont {Li}}, \bibinfo {author} {\bibfnamefont
  {Zhong}\ \bibnamefont {Wang}}, \ and\ \bibinfo {author} {\bibfnamefont
  {Faxian}\ \bibnamefont {Xiu}},\ }\bibfield  {title} {\enquote {\bibinfo
  {title} {Zeeman splitting and dynamical mass generation in {Dirac} semimetal
  {ZrTe}5},}\ }\href {\doibase 10.1038/ncomms12516} {\bibfield  {journal}
  {\bibinfo  {journal} {Nature Communications}\ }\textbf {\bibinfo {volume}
  {7}},\ \bibinfo {pages} {12516} (\bibinfo {year} {2016})}\BibitemShut
  {NoStop}%
\bibitem [{\citenamefont {Wang}\ \emph {et~al.}(2016)\citenamefont {Wang},
  \citenamefont {Li}, \citenamefont {Liu}, \citenamefont {Yan}, \citenamefont
  {Wang}, \citenamefont {Liu}, \citenamefont {Lin}, \citenamefont {Li},
  \citenamefont {Wang}, \citenamefont {Li}, \citenamefont {Mandrus},
  \citenamefont {Xie}, \citenamefont {Feng},\ and\ \citenamefont
  {Wang}}]{wang_chiral_2016}%
  \BibitemOpen
  \bibfield  {author} {\bibinfo {author} {\bibfnamefont {Huichao}\ \bibnamefont
  {Wang}}, \bibinfo {author} {\bibfnamefont {Chao-Kai}\ \bibnamefont {Li}},
  \bibinfo {author} {\bibfnamefont {Haiwen}\ \bibnamefont {Liu}}, \bibinfo
  {author} {\bibfnamefont {Jiaqiang}\ \bibnamefont {Yan}}, \bibinfo {author}
  {\bibfnamefont {Junfeng}\ \bibnamefont {Wang}}, \bibinfo {author}
  {\bibfnamefont {Jun}\ \bibnamefont {Liu}}, \bibinfo {author} {\bibfnamefont
  {Ziquan}\ \bibnamefont {Lin}}, \bibinfo {author} {\bibfnamefont {Yanan}\
  \bibnamefont {Li}}, \bibinfo {author} {\bibfnamefont {Yong}\ \bibnamefont
  {Wang}}, \bibinfo {author} {\bibfnamefont {Liang}\ \bibnamefont {Li}},
  \bibinfo {author} {\bibfnamefont {David}\ \bibnamefont {Mandrus}}, \bibinfo
  {author} {\bibfnamefont {X.~C.}\ \bibnamefont {Xie}}, \bibinfo {author}
  {\bibfnamefont {Ji}~\bibnamefont {Feng}}, \ and\ \bibinfo {author}
  {\bibfnamefont {Jian}\ \bibnamefont {Wang}},\ }\bibfield  {title} {\enquote
  {\bibinfo {title} {Chiral anomaly and ultrahigh mobility in crystalline
  {HfTe}$_5$},}\ }\href {\doibase 10.1103/PhysRevB.93.165127} {\bibfield
  {journal} {\bibinfo  {journal} {Physical Review B}\ }\textbf {\bibinfo
  {volume} {93}},\ \bibinfo {pages} {165127} (\bibinfo {year}
  {2016})}\BibitemShut {NoStop}%
\bibitem [{\citenamefont {Heikes}\ and\ \citenamefont
  {Ure}(1961)}]{heikes_thermoelectricity:_1961}%
  \BibitemOpen
  \bibfield  {author} {\bibinfo {author} {\bibfnamefont {Robert~R}\
  \bibnamefont {Heikes}}\ and\ \bibinfo {author} {\bibfnamefont {Roland~W}\
  \bibnamefont {Ure}},\ }\href@noop {} {\emph {\bibinfo {title}
  {Thermoelectricity: science and engineering}}}\ (\bibinfo  {publisher}
  {Interscience Publishers},\ \bibinfo {address} {New York},\ \bibinfo {year}
  {1961})\BibitemShut {NoStop}%
\bibitem [{\citenamefont {Jeon}\ \emph {et~al.}(2014)\citenamefont {Jeon},
  \citenamefont {Zhou}, \citenamefont {Gyenis}, \citenamefont {Feldman},
  \citenamefont {Kimchi}, \citenamefont {Potter}, \citenamefont {Gibson},
  \citenamefont {Cava}, \citenamefont {Vishwanath},\ and\ \citenamefont
  {Yazdani}}]{jeon_landau_2014}%
  \BibitemOpen
  \bibfield  {author} {\bibinfo {author} {\bibfnamefont {Sangjun}\ \bibnamefont
  {Jeon}}, \bibinfo {author} {\bibfnamefont {Brian~B.}\ \bibnamefont {Zhou}},
  \bibinfo {author} {\bibfnamefont {Andras}\ \bibnamefont {Gyenis}}, \bibinfo
  {author} {\bibfnamefont {Benjamin~E.}\ \bibnamefont {Feldman}}, \bibinfo
  {author} {\bibfnamefont {Itamar}\ \bibnamefont {Kimchi}}, \bibinfo {author}
  {\bibfnamefont {Andrew~C.}\ \bibnamefont {Potter}}, \bibinfo {author}
  {\bibfnamefont {Quinn~D.}\ \bibnamefont {Gibson}}, \bibinfo {author}
  {\bibfnamefont {Robert~J.}\ \bibnamefont {Cava}}, \bibinfo {author}
  {\bibfnamefont {Ashvin}\ \bibnamefont {Vishwanath}}, \ and\ \bibinfo {author}
  {\bibfnamefont {Ali}\ \bibnamefont {Yazdani}},\ }\bibfield  {title} {\enquote
  {\bibinfo {title} {Landau quantization and quasiparticle interference in the
  three-dimensional {Dirac} semimetal {Cd}3as2},}\ }\href {\doibase
  10.1038/nmat4023} {\bibfield  {journal} {\bibinfo  {journal} {Nature
  Materials}\ }\textbf {\bibinfo {volume} {13}},\ \bibinfo {pages} {851--856}
  (\bibinfo {year} {2014})}\BibitemShut {NoStop}%
\end{thebibliography}%

\end{document}